\documentclass[pra,twocolumn,superscriptaddress,showpacs,amsmath,amstex,amssymb,citeautoscript]{revtex4-1}

\usepackage[T1]{fontenc}
\usepackage{booktabs}
\usepackage[utf8]{inputenc}
\usepackage{amsmath}
\usepackage{amssymb}
\usepackage{bbm}
\usepackage{braket}
\usepackage{xcolor}
\usepackage{pifont}
\usepackage[mathscr]{euscript}
\usepackage[shortlabels]{enumitem}
\usepackage{graphicx}
\allowdisplaybreaks
\usepackage{float}
\usepackage{graphicx}
\usepackage{dsfont}
\usepackage{comment}
\usepackage[colorlinks=true]{hyperref}  
\hypersetup{
    bookmarks=true,         
    unicode=false,          
    pdftoolbar=true,        
    pdfmenubar=true,        
    pdffitwindow=false,     
    pdfstartview={FitH},    
    pdfsubject={},   
    pdfcreator={},   
    pdfproducer={}, 
    pdfkeywords={} {} {}, 
    pdfnewwindow=true,      
    colorlinks=true,       
    linkcolor=blue, 
    citecolor=blue,        
    filecolor=magenta,      
    urlcolor=blue           
}

\newcommand{\equref}[1]{Eq.~(\ref{#1})}

\newcommand{\figref}[1]{Fig.~\ref{#1}}
\newcommand{\refcite}[1]{Ref.~\onlinecite{#1}}

\definecolor{wrongultramarine}{rgb}{1,0.5,0}

\renewcommand{\vec}[1]{\boldsymbol{#1}}

\begin{document}

\title{Altermagnetic spin textures coupled to superconductors: \\ Domain wall spin-triplet superconductivity and supercurrent-induced torques}

\author{Yasir Dar}
\affiliation{Hearne Institute of Theoretical Physics, Department of Physics \& Astronomy, Louisiana State University, Baton Rouge LA 70803, USA}

\author{Mathias S.~Scheurer}
\email[Correspondence to: ]{mathias.scheurer@itp3.uni-stuttgart.de}
\affiliation{Institute for Theoretical Physics III, University of Stuttgart, 70550 Stuttgart, Germany}

\author{Constantin Schrade}
\email[Correspondence to: ]{cschrade@lsu.edu}
\affiliation{Hearne Institute of Theoretical Physics, Department of Physics \& Astronomy, Louisiana State University, Baton Rouge LA 70803, USA}

\begin{abstract}
Motivated by the absence of sizable stray fields and the recently discovered highly non-trivial impact of altermagnetic textures on itinerant electrons, we here study the form of Cooper pairs in spatially varying altermagnets coupled to conventional $s$-wave superconductors. As a consequence of the detrimental impact of altermagnetism on spin-singlet pairing and the local symmetry reduction caused by textures in the magnetic order parameter, we show that superconductivity predominantly impacts the regions between altermagnetic domains. Focusing on a planar radial domain wall for concreteness, we show that emergent Zeeman and spin-orbit fields create spatially separated triplet hotspots and transitions between nodal and fully gapped superconducting regions, whose structure is set by both the domain wall and the altermagnetic order parameter. We also identify a reciprocal effect, where a supercurrent generates a quasiparticle-mediated quadrupolar torque that inherits the symmetry of the altermagnetic order. Our results show that accounting for spatial inhomogeneities in the altermagnetic order parameter is essential for an understanding of the superconducting proximity effect and suggest that hybrid systems of altermagnetic textures and superconductors offer unique opportunities for local engineering of Cooper pairs and for detecting altermagnetic order.
\end{abstract}

\maketitle

\section{Introduction}
Large-scale spatial textures in the magnetic order parameter, such as domain walls, naturally occur in many magnetic materials \cite{hubert1998magnetic,Kumar2022May,Venkat2023Nov,Fert2017Jun} as a result of the competition of different interaction energies, complex dynamics during the formation of magnetic order, and the interplay with short-range inhomogeneities ``pinning’’ them. As such, spin textures are of fundamental importance for understanding the physics of magnets, both because they can crucially modify expectations based on theoretical considerations for the idealized homogeneous magnet and because they provide additional opportunities for rich physics and applications \cite{Fert2017Jun,Zhou2025Jan,Kumar2022May,Venkat2023Nov,parkin2008magnetic,fert2013skyrmions}. For instance, such spatial ($\vec{r}$) variations in the orientation $\vec{n}(\vec{r})$ of the magnetic order parameter have been shown to give rise to emergent electrodynamics for metallic ferromagnets \cite{Volovik1987,PhysRevLett.93.096806,PhysRevLett.102.067201,nam2009,PhysRevLett.98.246601,schulz2012,qxnw-8q4y} and antiferromagnets \cite{PhysRevB.86.245118,PhysRevB.91.144421,okabayashi2015theory,PhysRevB.93.180408}, where spatial gradients in $\vec{n}$ alter the effective hopping amplitudes of the itinerant electrons.

\begin{figure}[b!]
    \centering
    \includegraphics[width=1\linewidth]{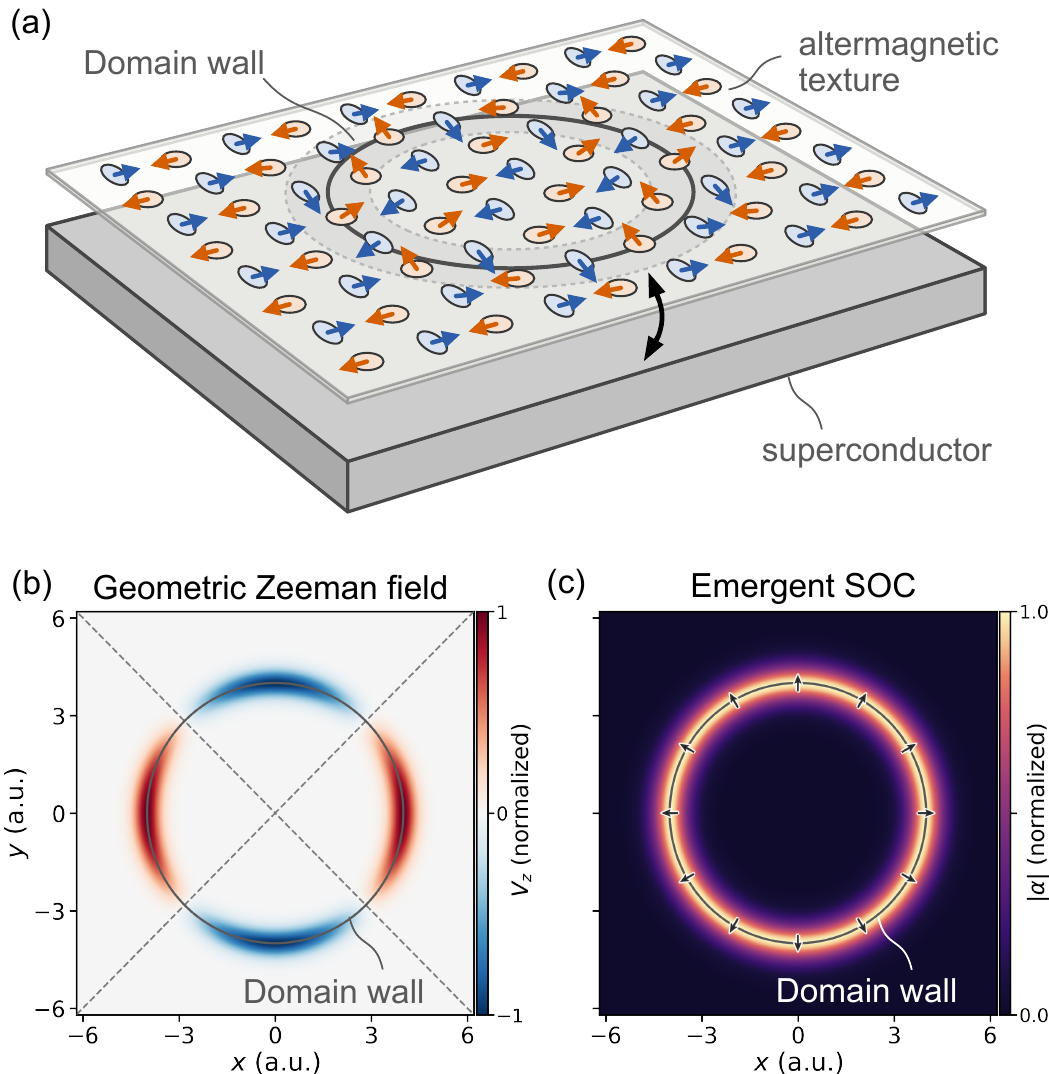}
\caption{
\textbf{Circular altermagnetic Néel domain wall proximitized by a superconductor.}
(a) Altermagnetic texture on a superconducting layer. Orange and blue ellipses denote the two sublattices ($\tau_z=\pm$). Arrows show the local in-plane Néel vector orientation. The circle marks the wall radius, $R$, and the shaded annulus marks the domain wall region where the Néel vector rotates.
(b) Geometric Zeeman field, $V_z(\boldsymbol{r})$, normalized by its maximum. The $d$-wave altermagnetic order gives a fourfold multipolar pattern.
(c) Emergent spin-orbit coupling, $\boldsymbol\alpha(\boldsymbol{r})$, normalized by its maximum. It is localized at the wall and points radially. Panel (a) is schematic and not to scale.
}
    \label{fig:1}
\end{figure}

Recently, altermagnets \cite{PhysRevX.12.040002,PhysRevX.12.040501} have been established as another class of magnets with favorable properties in the itinerant case: while symmetry guarantees that the net magnetization vanishes, the non-relativistic spin-degeneracy of the electronic bands is spontaneously split by the altermagnetic order parameter. Driven by their potential for spintronic applications, altermagnets have been actively studied, both theoretically and experimentally, in recent years \cite{PhysRevX.12.040501,Review3,magnetism5030017,PhysRevX.12.040002,RafaelReview}. However, although domain walls are known to be ubiquitous in altermagnetic materials, most works have focused on the homogeneous limit, $\vec{n}(\vec{r})=\vec{n}_0$. Apart from studies on the impact on the order parameter fields themselves \cite{DomainWallsSinova, PhysRevLett.133.196701,PhysRevLett.134.176401,PhysRevB.111.064422,2026arXiv260707789B} and the consequences of atomic-scale defects \cite{PhysRevB.111.174436,f6nc-vsnx,PhysRevB.110.205114,PhysRevB.111.035132,PhysRevB.110.165413,PhysRevB.111.L100502,PhysRevB.108.054510,2025arXiv251019943S}, the consequences of slowly varying altermagnetic textures on the electronic spectral properties have only very recently been addressed \cite{2026arXiv260214950M,2026arXiv260220236S}.

Following the nomenclature of \refcite{2026arXiv260220236S}, a coplanar texture in an altermagnet, like the circular domain wall shown in \figref{fig:1}(a), leads to two main effects, to leading order in the inverse exchange coupling constant: first, there is an emergent Zeeman field, $V_z$, which is unique to altermagnets, i.e., vanishes in the antiferromagnetic limit, and encodes the dominant orbital character of its order parameter. For instance, for a $d$-wave altermagnet, $V_z$ exhibits four sign changes upon encircling a domain wall, as illustrated in \figref{fig:1}(b). Second, the spatial variation in $\vec{n}$ induces a spin-orbit coupling term (which survives in the antiferromagnetic limit, in contrast to $V_z$); at fixed position $\vec{r}$, it is proportional to the projection of the momentum of the electrons on the vector $\vec{\alpha}(\vec{r})$, which is oriented perpendicular to the domain wall, see \figref{fig:1}(c).

Motivated by these findings, which reveal the highly non-trivial impact of textures in altermagnets on the Bloch Hamiltonian, we here explore how such inhomogeneities in the order parameter can be used to locally induce and control superconducting correlations and further study the inverse effect of supercurrent-driven control of altermagnetic textures. While our theory also applies to itinerant altermagnets that become superconducting at low enough temperatures, i.e., altermagnetism and superconductivity emerge spontaneously in the same material, an experimentally likely more flexible scenario is based on the proximity effect \cite{PhysRevB.111.L100502,2025arXiv250903774H,p79l-rty6,7h2t-wmyr,2026arXiv260103348S}: as shown in \figref{fig:1}(a), in a heterostructure of a superconductor and an altermagnet, the two phases influence each other via the superconducting proximity effect and exchange interactions.

In this work, we assume that the unperturbed superconductor is in a conventional spin-singlet pairing state. Since the altermagnet lifts the spinful Kramers degeneracy, it suppresses the singlet pairing correlations. However, spatial variations in its order parameter lead to $\vec{\alpha} \neq 0$, which locally induces triplet components that gap out the Fermi surfaces -- reminiscent of domain wall superconductivity discussed in ferromagnets \cite{PhysRevB.67.020503,Yang2004Nov,PhysRevB.111.174509}. In addition, the anisotropies around domain walls inherited from the orbital nature of the altermagnet lead to ``hotspots’’ of the admixed triplet component at certain positions on the domain wall, and the momentum dependence of the admixed equal-spin triplet component is shown to wind upon encircling the domain wall. Whether the local spectrum is fully gapped or has point nodes also depends on the position on the domain wall. Finally, we uncover a supercurrent-induced quadrupolar torque distorting the domain wall boundary, which is again associated with $V_z$ and comes on top of other forces, e.g., those moving the domain wall and discussed in \cite{2026arXiv260322243V}. More generally, our work generalizes the emergent field of superconducting altermagnets \cite{Fukaya_2025}, which---with the exception of atomic-scale disorder studies \cite{PhysRevB.111.174436,f6nc-vsnx,PhysRevB.111.L100502,PhysRevB.108.054510,2025arXiv251019943S}---primarily focused on the homogeneous limit. It further shows that altermagnets provide a very versatile alternative to ferromagnets \cite{Linder2015Apr,RevModPhys.77.1321} to locally engineer the Cooper-pair wave function; they not only suppress the detrimental stray fields but also allow additional effects, like $V_z$ and the associated triplet hotspots, opening up new paths for superconducting spintronics.

\section{Normal-state}
To define the theoretical modeling and the notation, we start with a concise derivation of the effective normal-state Hamiltonian for itinerant electrons moving through an altermagnetic texture, following \refcite{2026arXiv260220236S}.
Our starting point is the altermagnetic texture model
\begin{equation}
\hat{h}(\boldsymbol{r},\hat{\boldsymbol{p}})=
\varepsilon_{0,\hat{\boldsymbol{p}}}
+
t_{x,\hat{\boldsymbol{p}}}\tau_{x}
+
t_{z,\hat{\boldsymbol{p}}}\tau_{z}
+
J\tau_{z}\boldsymbol{n}(\boldsymbol{r})\cdot\boldsymbol{s}
-\mu
\label{Eq1}
\end{equation}
which generalizes the minimal models of uniform altermagnets \cite{PhysRevB.110.144412,PhysRevB.111.174436,839n-rckn} to a spatially varying Néel vector, 
$\boldsymbol{n}(\boldsymbol{r})$.
Here, $\boldsymbol{\tau}=(\tau_{x},\tau_{y},\tau_{z})$ acts in sublattice space,
$\boldsymbol{s}=(s_{x},s_{y},s_{z})$ in spin space,
and $\hat{\boldsymbol{p}}=-i\hbar(\partial_{x},\partial_{y})^{T}$.
The term $\varepsilon_{0,\hat{\boldsymbol{p}}}$ is sublattice independent,
while $t_{x,\hat{\boldsymbol{p}}}$ and $t_{z,\hat{\boldsymbol{p}}}$ describe inter-sublattice
and sublattice-antisymmetric hopping.
For concreteness, we focus on a $d$-wave texture with
$\varepsilon_{0,\hat{\boldsymbol{p}}}=\varrho_{0}(\hat{p}_{x}^{2}+\hat{p}_{y}^{2})$,
$t_{x,\hat{\boldsymbol{p}}}=\varrho_{x}+\varrho_{3}(\hat{p}_{x}^{2}+\hat{p}_{y}^{2})$,
and
$t_{z,\hat{\boldsymbol{p}}}=\varrho_{z}(\hat{p}_{x}^{2}-\hat{p}_{y}^{2})$.
For a uniform Néel vector, Eq.~\eqref{Eq1} reproduces the usual $d$-wave altermagnetic band splitting.

It will be convenient to work in a rotating frame where the local spin quantization axis
aligns with the direction of the Néel vector. 
We therefore choose a unitary transformation 
$U(\boldsymbol{r})$ 
such that
$U^\dagger[\boldsymbol{n}(\boldsymbol{r})\cdot\boldsymbol{s}]U=s_z$.
The spatial dependence of the texture in this rotating frame enters the Hamiltonian through a shift of the momentum operator by a spin-dependent gauge field, 
$
\hat{\boldsymbol\pi}
\equiv 
U^\dagger\hat{\boldsymbol p}
U=\hat{\boldsymbol p}-\boldsymbol{A}(\boldsymbol{r})
$ 
with 
$A_i(\boldsymbol{r})=\boldsymbol{\alpha}_i(\boldsymbol{r})\cdot\boldsymbol{s}$. 
The full rotating frame Hamiltonian reads
\begin{equation}
\hat{h}_{\text{rot}}(\boldsymbol{r},\hat{\boldsymbol{\pi}})=
\varepsilon_{0,\hat{\boldsymbol{\pi}}}
+
t_{x,\hat{\boldsymbol{\pi}}}\tau_{x}
+
t_{z,\hat{\boldsymbol{\pi}}}\tau_{z}
+
J\tau_{z}s_{z}
\label{Eq2}
\end{equation}

Let us now take $J>0$ as the largest energy scale and project Eq.~\eqref{Eq2} onto the low-energy subspace with $S\equiv\tau_z s_z=-1$ with projector $P_-=(1-S)/2$. From this projection, we will obtain an effective Hamiltonian with emergent fields that act on the low-energy spin. 
We split the gauge field into longitudinal and transverse parts, 
$A_i=A_i^\parallel+A_i^\perp$. 
The longitudinal part,
$A_i^\parallel=\alpha_i^z(\boldsymbol{r})s_z$,
acts within the low-energy subspace, 
while the transverse part,
$A_i^\perp=\alpha_i^x s_x+\alpha_i^y s_y$, 
couples the low- and high-energy subspaces. 
We further define the longitudinally shifted momentum 
$\hat\Pi_i=\hat p_i-A_i^\parallel(\boldsymbol{r})$, 
so that $\hat\pi_i=\hat\Pi_i-A_i^\perp(\boldsymbol{r})$. 
The projection of Eq.~\eqref{Eq2} onto the $S=-1$ subspace gives
\begin{align}
\hat{h}_{\text{proj}}(\boldsymbol{r},\hat{\boldsymbol{\Pi}})
&=
\varepsilon_{0,\hat{\boldsymbol{\Pi}}}
+
t_{z,\hat{\boldsymbol{\Pi}}}\sigma_{z}
+
\varrho_{0}V_{0}(\boldsymbol{r})
+
\varrho_{z}V_{z}(\boldsymbol{r})\sigma_{z}
\nonumber
\\
&\hspace{1pt}
+
\hat{h}_{\text{SOC}}(\boldsymbol{r},\hat{\boldsymbol{\Pi}})
\label{Eq3}
\end{align}
Here,
$\sigma_{x}=P_{-}\tau_{x}s_{x}P_{-}$,
$\sigma_{y}=-P_{-}\tau_{x}s_{y}P_{-}$,
and
$\sigma_{z}=P_{-}\tau_{z}P_{-}$
are the Pauli matrices in the low-energy subspace; due to spin-sublattice locking, one can think of $\sigma_j$ as acting on either spin or sublattice degrees of freedom. 

Two comments about the low-energy effective Hamiltonian in Eq.~\eqref{Eq3} are in order: First, the itinerant electrons are subject to a scalar potential, $V_0(\boldsymbol{r})$, and an effective Zeeman field, $V_z(\boldsymbol{r})$. 
These fields are given by 
\begin{equation}
\begin{split}
V_{0}(\boldsymbol{r})
&=
\hbar^2
\delta^{ij}g_{ij}(\boldsymbol{r}),
\\
V_{z}(\boldsymbol{r})
&=
\hbar^2
\eta^{ij}g_{ij}(\boldsymbol{r}),
\label{Eq4}
\end{split}
\end{equation}
where $\eta^{ij}=\text{diag}(1,-1)^{ij}$  and
$
g_{ij}(\boldsymbol{r})
=
\partial_i\boldsymbol{n}(\boldsymbol{r})\cdot\partial_j\boldsymbol{n}(\boldsymbol{r})/4
$.
Notably, $V_z(\boldsymbol{r})$
acts as a \textit{spatially varying spin splitting} in the low-energy subspace. 
It is unique to the altermagnet and vanishes in the antiferromagnetic limit, $\varrho_z=0$. The fact that $\vec{k}^T\eta \vec{k} = k_x^2-k_y^2$ encodes the orbital character of the altermagnet and $g$ is the real-space quantum metric associated with the texture Hamiltonian 
$\boldsymbol{n}(\boldsymbol{r})\cdot\boldsymbol{s}$ underscores the geometric nature of $V_0$ and $V_z$.

Second, the itinerant electrons are also subject to an emergent spin-orbit coupling, generated by the transverse gauge field,
\begin{equation}
\hat{h}_{\text{SOC}}(\boldsymbol{r},\hat{\boldsymbol{p}})
=
-\varrho_{3}
\sum_{i=x,y}
\left[
\{\hat{p}_{i},\alpha_{i}^{x}(\boldsymbol{r})\}\sigma_{x}
-
\{\hat{p}_{i},\alpha_{i}^{y}(\boldsymbol{r})\}\sigma_{y}
\right].
\end{equation}
This term provides a \textit{spatially varying spin mixing}.
It can also be written as
$
\hat{h}_{\text{SOC}}
=
-\varrho_{3}\sum_{i=x,y}P_{-} \tau_{x} \{\hat{p}_{i},A_{i}^{\perp}\}P_{-},
$
from which it follows that
$
\hat{h}_{\text{SOC}}(\boldsymbol{r},\hat{\boldsymbol{p}})
=
\hat{h}_{\text{SOC}}(\boldsymbol{r},\hat{\boldsymbol{\Pi}})
$
since $\{A_{i}^{\parallel},A_{i}^{\perp}\}=0$. 

A natural question is: How do these emergent fields affect the Cooper pair wavefunction upon proximitizing the altermagnetic texture and a superconductor, as illustrated schematically in \figref{fig:1}(a)?

\section{Superconducting state}
To address this question, let us assume for concreteness that the superconductor is a conventional, $s$-wave spin-singlet state, inducing a momentum-independent spin-singlet pairing potential $\hat{\Delta}(\boldsymbol{r}) \propto i s_y$ in the altermagnet.
Fermionic antisymmetry, 
$\hat\Delta=-\hat\Delta^T$, 
allows the sublattice matrix multiplying $i s_y$ 
to be $\tau_0$, $\tau_x$, or $\tau_z$. 
The intra-sublattice singlets, $\tau_0 i s_y$ and $\tau_z i s_y$, couple $S=-1$ to $S=+1$ 
and thus vanish under projection onto the low-energy subspace. 
The inter-sublattice singlet, $\tau_x i s_y$, pairs the low-energy states directly and survives the projection onto the low-energy subspace. Hence, we will focus on
\begin{equation}
\hat{\Delta}(\boldsymbol{r})
=
\Delta(\boldsymbol{r})\,\tau_xi s_y
\end{equation}
in the following, obeying $P_-\hat\Delta P_-=\Delta(\boldsymbol{r})(-i\sigma_y)$ in the low-energy subspace. The effective Bogoliubov-de Gennes (BdG) Hamiltonian then takes the form
\begin{equation}
\hat{\mathcal H}_{\text{BdG}}(\boldsymbol{r},\hat{\boldsymbol\Pi})
=
\begin{pmatrix}
\hat{h}_{\text{proj}}(\boldsymbol{r},\hat{\boldsymbol\Pi})
&
\Delta(\boldsymbol{r})\,(-i\sigma_y)
\\
-\Delta^*(\boldsymbol{r})\,(-i\sigma_y)
&
-[\hat{h}_{\text{proj}}(\boldsymbol{r},-\hat{\boldsymbol\Pi})]^T
\end{pmatrix}.
\label{Eq7}
\end{equation}
This effective Hamiltonian shows that, within the low-energy subspace, the texture is subject to
an effective $s$-wave pairing, in addition to the texture-induced fields. All texture-induced fields are spatially varying, and $V_z(\boldsymbol{r})$ retains the $d$-wave structure of the altermagnetic order through the form factor, $\eta^{ij}$. Our expectation is that the texture and the altermagnetic form factor control the local spin dependence of the induced pairing correlations.

\section{Nambu-space Green's function}
The pairing correlations are given by the off-diagonal components of the Nambu-space Green's function. We will now discuss how to compute the Nambu-space Green's function in a local, semiclassical approximation.

Our focus will be, for concreteness, on a planar radial domain wall, 
$\boldsymbol{n}(\boldsymbol{r})=(\cos\phi(r),\sin\phi(r),0)$, 
with 
$\phi(r)=(\pi/2)\tanh[(r-R_0)/w]$. 
Here, 
$R_0$ is the wall radius, 
$w$ the wall width, and
$(r,\chi)$ are polar coordinates with $\boldsymbol{r}=(r\cos\chi,r\sin\chi)$.

For this radial domain wall texture, the projected normal-state Hamiltonian of Eq.~\eqref{Eq3} simplifies greatly. 
The longitudinal gauge field can be removed by the unitary, 
$W(\boldsymbol{r}) \hat\Pi_{x,y}W^\dagger(\boldsymbol{r})=\hat p_{x,y}$ 
with
$W(\boldsymbol{r})=e^{-i\phi(r)\sigma_z/2}$. 
The emergent spin-orbit coupling simplifies to
$
W\hat{h}^{(0)}_{\text{SOC}}W^\dagger
=
(1/2)\{\hat{\boldsymbol p},\boldsymbol{\alpha}(r)\}
\sigma_x,
$
where
$
\boldsymbol{\alpha}(r) 
=
\hbar \varrho_3 \phi'(r)\hat{\boldsymbol r}
$
with the radial unit vector 
$
\hat{\boldsymbol r}
$; see \figref{fig:1}(c).
The effective normal-state Hamiltonian then takes on the form
\begin{equation}
\hat{h}_{\text{proj}}(\boldsymbol r,\hat{\boldsymbol{p}})
=
\xi(\boldsymbol{r},\hat{\boldsymbol{p}})
\sigma_0
+
b_{z}(\boldsymbol{r},\hat{\boldsymbol{p}})
\sigma_z
+
\frac{1}{2}
\{\hat{\boldsymbol p},\boldsymbol{\alpha}(r)\}
\sigma_x.
\label{Eq8}
\end{equation}
Here, the spin-independent term is
$\xi(\boldsymbol{r},\hat{\boldsymbol{p}})
=
\varrho_0 \hat{\boldsymbol{p}}^2 + \varrho_0 V_0(r)-\mu
$
with
$
V_0(r)=\frac{\hbar^2}{4}\phi'(r)^2
$.
The spin-splitting term is
$
b_{z}(\boldsymbol{r},\hat{\boldsymbol{p}})
=
\varrho_z(\hat{\boldsymbol{p}}\cdot\eta\hat{\boldsymbol{p}})+\varrho_z V_z(r,\chi)
$
with
$
V_z(r,\chi)=\frac{\hbar^2}{4}\phi'(r)^2 \hat{\boldsymbol{r}}\cdot\eta\hat{\boldsymbol{r}}
=\frac{\hbar^2}{4}\phi'(r)^2 \cos2\chi
$
and
$
\hat{\boldsymbol{r}} = (\cos\chi,\sin\chi).
$
A consequence of  Eq.~\eqref{Eq8} is that the BdG Hamiltonian for the planar domain wall can be written in terms of the usual momentum operator, 
$\hat{\mathcal{H}}_{\text{BdG}}(\boldsymbol{r},\hat{\boldsymbol{\Pi}})
=
\hat{\mathcal{H}}_{\text{BdG}}(\boldsymbol{r},\hat{\boldsymbol{p}})$.

Let us now discuss the Nambu-space Green's function. It is defined as the solution of the Gor'kov equation, 
$
(
i\omega_{n}
-
\hat{\mathcal{H}}_{\text{BdG}}\left(\boldsymbol{r}_{1},\hat{\boldsymbol{p}}_{1}\right)
)
\mathcal{G}(\boldsymbol{r}_{1},\boldsymbol{r}_{2}; i\omega_{n})
=
\delta\left(\boldsymbol{r}_{1}-\boldsymbol{r}_{2}\right)
$
where $\omega_n$ are the fermionic Matsubara frequencies. 
We will rewrite the Gor'kov equation in a ``phase-space form'', 
which has been previously used in the quasiclassical theory of superconductivity~\cite{nakamura}.
We therefore adopt a position-space representation of operators, 
$\mathcal{O}(\boldsymbol{r}_1,\boldsymbol{r}_2)
\equiv
\langle \boldsymbol{r}_1|\hat{\mathcal{O}}|\boldsymbol{r}_2\rangle$. 
We then introduce center-of-mass and relative coordinates,
$
\boldsymbol{R}
=
(\boldsymbol{r}_1 + \boldsymbol{r}_2)/2
$
and
$
\boldsymbol{\rho}
=
\boldsymbol{r}_1-\boldsymbol{r}_2,
$
and perform the ``Wigner transform'',  
$
\mathcal{O}(\boldsymbol{R}, \boldsymbol{p})
=
\int d^2\rho\;
e^{-\frac{i}{\hbar} \boldsymbol{p} \cdot \boldsymbol{\rho}}
\mathcal{O} \left(
\boldsymbol{R} + \frac{\boldsymbol{\rho}}{2},
\boldsymbol{R} - \frac{\boldsymbol{\rho}}{2}
\right).
$
Here, $\boldsymbol{p}$ is the relative momentum. 
The Gor'kov equation then takes on the form, 
\begin{equation}
\left(
i \omega_n
-
\mathcal{H}_{\text{BdG}}(\boldsymbol{R}, \boldsymbol{p})
\right)
\star
\mathcal{G}(\boldsymbol{R}, \boldsymbol{p}; i\omega_n)
=
1.
\label{Eq9}
\end{equation}
Here, 
$\mathcal{H}_{\text{BdG}}(\boldsymbol{R}, \boldsymbol{p})$ 
and
$\mathcal{G}(\boldsymbol{R}, \boldsymbol{p}; i\omega_n)$
are matrix-valued functions on the classical phase space, $(\boldsymbol{R},\boldsymbol{p})$.
Quantum corrections are encoded in the ``Moyal product'', 
$
\mathcal{O}_1(\boldsymbol{R}, \boldsymbol{p}) \star \mathcal{O}_2(\boldsymbol{R}, \boldsymbol{p})
=
\mathcal{O}_1(\boldsymbol{R}, \boldsymbol{p})\,
\exp[
\frac{i \hbar}{2}
(
\overleftarrow{\nabla}_{\boldsymbol{R}} \cdot \overrightarrow{\nabla}_{\boldsymbol{p}}
-
\overleftarrow{\nabla}_{\boldsymbol{p}} \cdot \overrightarrow{\nabla}_{\boldsymbol{R}}
)
]
\mathcal{O}_2(\boldsymbol{R}, \boldsymbol{p}) 
$. 
This rewrite of the Gor'kov equation will be the starting point for the local, semiclassical approximation.

We now assume that the texture-induced fields, 
$V_{0}(\boldsymbol{r})$,
$V_{z}(\boldsymbol{r})$,
and
$\boldsymbol{\alpha}(\boldsymbol{r})$,
vary slowly on the scale of the Fermi wavelength and the superconducting coherence length. 
Under this assumption, the local, semiclassical approximation involves retaining only 
the lowest-order term in the expansion of the Moyal product, 
$
\mathcal{O}_1(\boldsymbol{R}, \boldsymbol{p}) \star \mathcal{O}_2(\boldsymbol{R}, \boldsymbol{p})
\approx
\mathcal{O}_1(\boldsymbol{R}, \boldsymbol{p}) \mathcal{O}_2(\boldsymbol{R}, \boldsymbol{p})
$. 
Interestingly, since the texture-induced fields are local functions in phase space,
this approximation already captures the leading order at which texture 
effects enter the Nambu-space Green's function. 
Specifically, we approximate the Nambu-space Green's function as,
\begin{equation}
\mathcal{G}(\boldsymbol{R},\boldsymbol{p};i\omega_{n})
\approx
\big[
i\omega_{n}-\mathcal{H}_{\text{BdG}}(\boldsymbol{R},\boldsymbol{p})
\big]^{-1}.
\label{Eq10}
\end{equation}
We will now use this approximate form to address the question on the spin structure of the pairing correlations and the quasiparticle spectrum.

\section{Quasiparticle spectrum}
We first discuss the local quasiparticle spectrum, because it will provide us with an intuition for the discussion on the pairing correlations. The local quasiparticle energies of 
$
\mathcal{H}_{\text{BdG}}(\boldsymbol{R},\boldsymbol{p})
$ 
are,
\begin{equation}
\begin{split}
E_{\pm}(\boldsymbol{p}) 
&=
[
\Delta^{2}
+
\xi(\boldsymbol{p})^{2}
+
\alpha_{\boldsymbol{p}}^{2}
+
b_{z}(\boldsymbol{p})^{2}
\\
&\quad\pm 
2\sqrt{
\xi(\boldsymbol{p})^{2}
[
\alpha_{\boldsymbol{p}}^{2}
+
b_{z}(\boldsymbol{p})^{2}]
+
\Delta^{2} 
b_{z}(\boldsymbol{p})^{2}}
]^{1/2}    
\end{split}    
\end{equation}
Here, we omitted the position argument, since we will work at fixed $\boldsymbol R$. 
The factor $\alpha_{\boldsymbol{p}}(\boldsymbol{R})\equiv \boldsymbol{p}\cdot \boldsymbol{\alpha}(R)$ with $R=|\boldsymbol{R}|$ arises from the emergent spin-orbit coupling.
For the radial domain wall,
$\alpha_{\boldsymbol{p}}(\boldsymbol{R})=\hbar\varrho_{3} \phi'(R)p_{R}$, 
where 
$p_{R}=\boldsymbol{p}\cdot \hat{\boldsymbol{r}}$ is the radial momentum.

To describe the effective low-energy pairing, we now move to the helicity basis, defined as the eigenbasis of the local normal-state Hamiltonian.  
The normal-state Hamiltonian is
$
h_{\text{proj}}(\boldsymbol{p})
=
\xi(\boldsymbol{p})\sigma_{0}
+
\alpha_{\boldsymbol{p}}\sigma_{x}
+
b_{z}(\boldsymbol{p})\sigma_{z}
$.
Its helicity eigenstates satisfy
$
h_{\text{proj}}(\boldsymbol{p})
\lvert s,\boldsymbol{p}\rangle
=
\left[
\xi(\boldsymbol{p})
+
s\lambda_{\boldsymbol{p}}
\right]
\lvert s,\boldsymbol{p}\rangle
$
with $s=\pm$ and
$
\lambda_{\boldsymbol{p}}
=
(
b_{z}(\boldsymbol{p})^{2}
+
\alpha_{\boldsymbol{p}}^{2}
)^{1/2}.
$
Equivalently, the unitary matrix, $U_{\boldsymbol{p}}$, whose columns are the helicity eigenstates, obeys,
$
U_{\boldsymbol{p}}^{\dagger}
h_{\text{proj}}(\boldsymbol{p})
U_{\boldsymbol{p}}
=
\xi(\boldsymbol{p})\sigma_{0}
+
\lambda_{\boldsymbol{p}}\sigma_{z}.
$
An explicit choice of eigenstates is 
$
\lvert +,\boldsymbol{p}\rangle
=
(
\cos(\theta_{\boldsymbol{p}}/2),
\sin(\theta_{\boldsymbol{p}}/2)
)^{T}
$
and 
$
\lvert -,\boldsymbol{p}\rangle
=
(-\sin(\theta_{\boldsymbol{p}}/2),
\cos(\theta_{\boldsymbol{p}}/2)
)^{T}
$ 
with
$
\cos\theta_{\boldsymbol{p}}=b_{z}/\lambda_{\boldsymbol{p}}
$
and
$
\sin\theta_{\boldsymbol{p}}=\alpha_{\boldsymbol{p}}/\lambda_{\boldsymbol{p}}
$.

\begin{figure}[!t]
    \centering
    \includegraphics[width=1\linewidth]{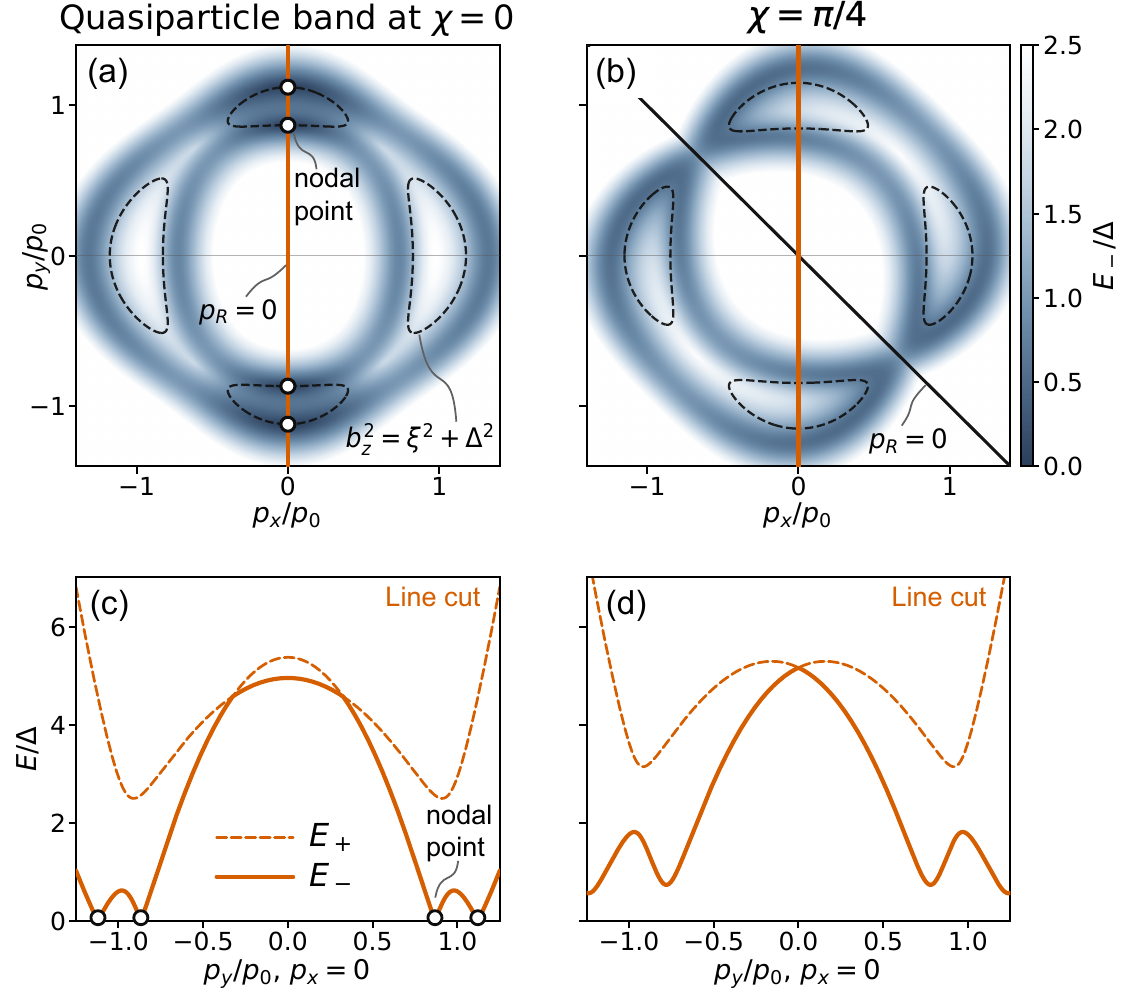}
\caption{
\textbf{Local quasiparticle spectrum for different points along a proximitized circular altermagnetic domain wall.}
(a,b) Lower positive-energy quasiparticle band,
$E_-(\boldsymbol R,\boldsymbol p)/\Delta$,
evaluated on the radial domain wall at the wall positions
$\chi=0$ and $\chi=\pi/4$.
Dashed contours indicate the condition $b_z^2=\xi^2+\Delta^2$.
A nodal point additionally requires
$p_R=0$, so that the emergent spin-orbit coupling vanishes. 
(c,d) At $\chi=0$, four nodal points appear in the quasiparticle spectrum. 
At $\chi=\pi/4$, the quasiparticle spectrum is fully gapped. The momentum normalization is
$p_0=\sqrt{\mu/\varrho_0}$. 
}
    \label{fig2}
\end{figure}

We next transform the superconducting order parameter to the helicity basis, 
$
\Delta_{\text{hel}}(\boldsymbol{p})
=
U_{\boldsymbol{p}}^{\dagger}
\Delta(\boldsymbol{p})
U_{-\boldsymbol{p}}^{*}.
$
We find the form 
\begin{equation}
\Delta_{\text{hel}}(\boldsymbol{p})
=
\frac{\Delta}{\lambda_{\boldsymbol{p}}}
\begin{pmatrix}
\alpha_{\boldsymbol{p}} & -b_{z} \\
b_{z} & \alpha_{\boldsymbol{p}}
\label{effectivepairing}
\end{pmatrix}.
\end{equation}
 The diagonal entries describe pairing between states in the same helicity band, $|+,\boldsymbol{p}\rangle\leftrightarrow|+,-\boldsymbol{p}\rangle$ 
and 
$|-,\boldsymbol{p}\rangle
\leftrightarrow
|-,-\boldsymbol{p}\rangle
$. 
By contrast, the off-diagonal terms pair states in different helicity bands,
$
|+,\boldsymbol{p}\rangle
\leftrightarrow
|-,-\boldsymbol{p}\rangle
$.

Suppose that, at a given momentum $\vec{p}$, only one helicity band, $s$, lies near the Fermi level, while the other is separated by an energy large compared with the induced pairing scale. 
We then neglect the interband pairing and retain only the pairing within band $s$. 
The resulting low-energy BdG Hamiltonian is
\begin{equation}
\mathcal{H}^{\text{(eff)}}_{\text{BdG},s}(\boldsymbol{p})
=
\begin{pmatrix}
\xi + s\lambda_{\boldsymbol{p}}
&
\Delta \alpha_{\boldsymbol{p}}/\lambda_{\boldsymbol{p}}
\\
\Delta \alpha_{\boldsymbol{p}}/\lambda_{\boldsymbol{p}}
&
-\xi - s\lambda_{\boldsymbol{p}}
\label{effectivehamiltonian}
\end{pmatrix}.    
\end{equation}
Its spectrum is
$
E_s^2
=
(\xi+s\lambda_{\boldsymbol{p}})^2
+
\Delta^2\alpha_{\boldsymbol{p}}^2/\lambda_{\boldsymbol{p}}^2
$. 
The normal-state dispersion, 
$
\xi(\boldsymbol{p})+s\lambda_{\boldsymbol{p}},
$
is even in momentum, whereas the intraband pairing amplitude is odd,
$\alpha_{-\boldsymbol{p}}=-\alpha_{\boldsymbol{p}}$
and
$
\lambda_{\boldsymbol{p}}=\lambda_{-\boldsymbol{p}}
$. 
The projected theory therefore describes an odd-parity superconductor within each nondegenerate helicity species. 

We can further understand the spin structure of induced pariring by transforming the superconducting order parameter from the helicity basis back to the $\sigma_z$ basis and decomposing it into singlet and triplet components as $[\psi(\boldsymbol p)\sigma_0+\boldsymbol d(\boldsymbol p)\cdot\boldsymbol\sigma]i\sigma_y$.
In this case, we find 
\begin{equation}
    \psi(\boldsymbol p) = \frac{\Delta \alpha^2_{\vec{p}}}{2 \lambda_{\vec{p}}^2}, \,\,\, \vec{d}(\boldsymbol p) = -\frac{\Delta \alpha_{\vec{p}}}{2 \lambda_{\vec{p}}}(\pm 1 , i b_z(\vec{p})/\lambda_{\vec{p}} ,0 )^T. \label{SingletTriplParts}
\end{equation}
Thus, in the limit of strong altermagnetism, we obtain predominantly nonunitary equal-spin triplet pairing, with
$\vec{d}=(\mp 1,i,0)^T\Delta\alpha/(2b_z)+\mathcal{O}(\alpha^2/b_z^2)$
and
$\psi=0+\mathcal{O}(\alpha^2/b_z^2)$.
The two signs correspond to the two helicity bands, which predominantly support the $\uparrow\uparrow$ and $\downarrow\downarrow$ pairing channels, respectively.
As can also be seen in \equref{SingletTriplParts}, the low-energy pairing correlations are only non-zero if $\alpha_{\vec{q}}\neq 0$ and, thus, only sizeable in the vicinity of the domain wall. Given the dominance of the triplet component, altermagnetic domain wall systems therefore realize \textit{domain wall spin-triplet superconductivity}.

In addition to studying the low-energy superconducting order parameter, we can also determine when the full spectrum 
is gapped and when nodal points arise. From 
$
E_+^2E_-^2
=
(
\xi^2+\Delta^2-\lambda_{\boldsymbol p}^2
)^2
+
4
\Delta^2\alpha_{\boldsymbol p}^2
$, 
we see that nodes only arise if simultaneously $\alpha_{\boldsymbol p}=0$ and $b_z^2=\xi^2+\Delta^2$. This behavior is illustrated in \figref{fig2}, where the momenta obeying these two constraints are shown as solid and dashed black lines, respectively. We can see that, depending on the position on the domain wall, which determines the orientation of the line with $\alpha_{\boldsymbol p}=0$, the superconducting spectrum either exhibits nodal points at the intersection of the two lines, see \figref{fig2}(a,c), or is fully gapped (no intersection of the solid and dashed lines), see \figref{fig2}(b,d). We emphasize that this directional dependence of the type of excitation spectrum is crucially connected to the altermagnetic nature of the system since $b_z^2=\xi^2+\Delta^2$ has no solutions for $b_z=0$, making the system generically fully gapped in the antiferromagnetic limit.

\begin{figure}[!t]
    \centering
    \includegraphics[width=1\linewidth]{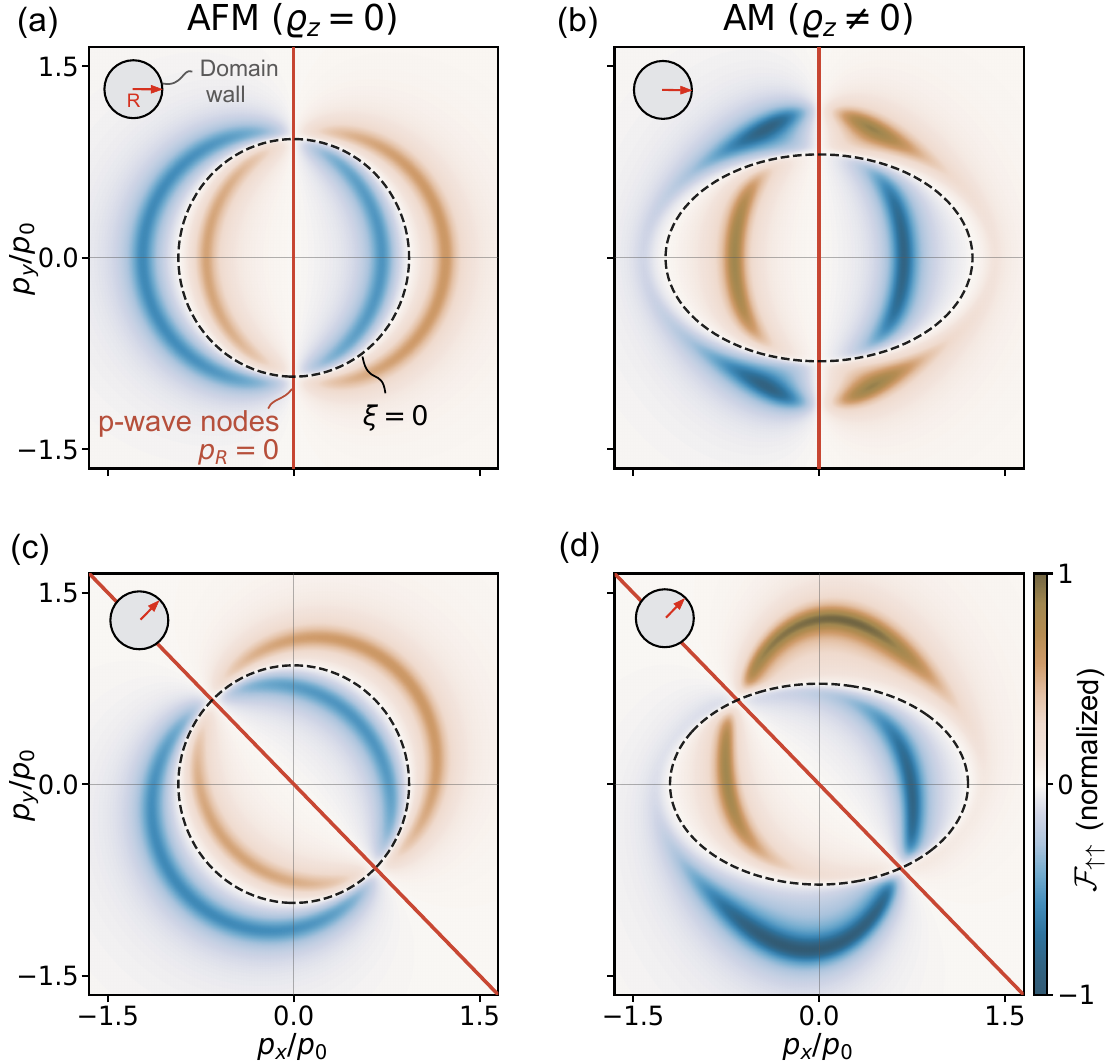}
\caption{
\textbf{Radial altermagnetic domain wall generates spin-polarized triplet correlations.}
Equal-spin triplet correlator,
$\mathcal{F}_{\uparrow\uparrow}(\boldsymbol{R},\boldsymbol{p})$,
for a proximitized radial altermagnetic domain wall normalized to its maximum. 
Insets indicate the position, $\boldsymbol{R}$, on the wall.
(a,c) Antiferromagnetic limit, $\varrho_z=0$. 
The emergent spin-orbit coupling generates a $p$-wave triplet component with nodes at 
$p_R=\boldsymbol{p}\cdot\hat{\boldsymbol{R}}=0$ (red line). Additional nodes arise at $\xi(\boldsymbol{R},\boldsymbol{p})=0$ (dashed line).
(b,d) Altermagnetic case, $\varrho_z\neq0$. The $p$-wave triplet component is now modulated by the $d$-wave altermagnetic form factor. Hence, the additional nodes arise at $\xi(\boldsymbol{R},\boldsymbol{p})-b_z(\boldsymbol{R},\boldsymbol{p})=0$.
}
    \label{fig:2}
\end{figure}

\section{Pairing correlations}
Having discussed the local quasiparticle spectrum and the effective low-energy BdG Hamiltonian, we will next turn to the induced pairing correlations. To this end, we study the anomalous electron-hole block of the Nambu-space Green's function. 
It reads
$
F(\boldsymbol R,\boldsymbol p;i\omega_n)
\equiv
\langle e|\mathcal G(\boldsymbol R,\boldsymbol p;i\omega_n)|h\rangle
$
where 
$|e\rangle=(1,0)^T$
and
$
|h\rangle=(0,1)^T
$
are vectors in Nambu space.
We focus on the equal-time pairing correlator, 
$
\mathcal F_{\sigma\sigma'}(\boldsymbol R,\boldsymbol p)
\equiv
T\sum_{\omega_n}
F_{\sigma\sigma'}(\boldsymbol R,\boldsymbol p;i\omega_n).
$
Evaluating Eq.~\eqref{Eq10} and performing the Matsubara sum at zero temperature gives,
\begin{equation}
\mathcal F_{\sigma\sigma}(\boldsymbol R,\boldsymbol p)
=
\frac{\Delta(\boldsymbol R)\alpha_{\boldsymbol p}(\boldsymbol R)}
{
D(\boldsymbol R,\boldsymbol p)
}
\left[
\sigma\xi(\boldsymbol R,\boldsymbol p)
-
b_z(\boldsymbol R,\boldsymbol p)
\right],
\label{Eq11}
\end{equation}
where $\sigma=+1$ for $\uparrow\uparrow$ pairs and $\sigma=-1$ for $\downarrow\downarrow$ pairs. 
We remark that, in this context, $\uparrow$ and $\downarrow$ are the labels for eigenstates of $\sigma_z$ in the local rotating frame, not in the fixed laboratory frame of the system. 
In the laboratory frame, $\uparrow$ would refer to a local spin orientation that is antiparallel to $\boldsymbol{n}(\boldsymbol{R})$, whereas 
$\downarrow$ would refer to a local spin orientation that is parallel to $\boldsymbol{n}(\boldsymbol{R})$. 
For our radial domain wall texture, we have $\boldsymbol n(r\ll R_0)=-\hat{\boldsymbol y}$, $\boldsymbol n(R_0)=+\hat{\boldsymbol x}$, and $\boldsymbol n(r\gg R_0)=+\hat{\boldsymbol y}$. 
As a result, the spin polarization of the Cooper pairs will reverse upon traversing the wall. 
However, since $\boldsymbol{n}(\boldsymbol{R})=\boldsymbol{n}(R)$ there is no azimuthal winding of the spin direction upon encircling the wall at a fixed radius, $R$.

Equation~\eqref{Eq11} is one of our main results and 
several comments are in order:

First, we note that equal-spin triplets are generated only where the texture produces a nonzero emergent spin-orbit coupling. 
For the radial domain wall,
$\alpha_{\boldsymbol p}(\boldsymbol R)=\hbar\varrho_3 \phi'(R)p_R$, 
where 
$p_R=\boldsymbol p\cdot \hat{\boldsymbol r}$ and $R=|\boldsymbol R|$. 
Triplet generation is therefore localized at the wall, where $\phi'(R)\neq0$. 
It is also momentum selective. 
We see that tangential momenta with $p_R=0$ give no conversion.
Thus, the equal-spin triplet correlator has nodes along $p_R=0$. 
Because the radial direction, $\hat{\boldsymbol r}$, changes around the wall, these nodes rotate when moving to different
wall positions, as we show with red lines in \figref{fig:2}.

Second, the factor in brackets of Eq.~\eqref{Eq11} contains a first contribution $\propto\xi(\boldsymbol R,\boldsymbol p)$.
This contribution is present even when the altermagnetic splitting is absent, $\varrho_z=0$.
In this situation, it produces nodes on the circle 
$\varrho_0 \boldsymbol{p}^2=\mu-\varrho_0 V_0(R)$, assuming that the right-hand side is positive, see dashed circles in \figref{fig:2}(a,c).

Third, the second contribution $\propto b_z(\boldsymbol R,\boldsymbol p)$ in Eq.~\eqref{Eq11} is $\propto\varrho_z$ and, thus, unique to the altermagnet. 
It changes the node condition to 
$\sigma\xi(\boldsymbol R,\boldsymbol p)-b_z(\boldsymbol R,\boldsymbol p)=0$, which evaluates to $(\varrho_0-\sigma\varrho_z)p_x^2+(\varrho_0+\sigma\varrho_z)p_y^2=\mu-\varrho_0V_0(R)+\sigma\varrho_z V_z(R,\chi)$. For $\varrho_0>|\varrho_z|$ and a positive right-hand side, this equation describes a spin-dependent ellipse in momentum space. Thus, the altermagnet
creates an anisotropy on the pairing correlations. This anisotropy will lead to a spatially nonuniform 
distribution of spin-polarized triplet correlations around the wall, as is visible in \figref{fig:2}(b,d). We will refer to the associated points on the domain wall with maximum net triplet contributions as ``triplet hotspots''.

To describe the formation of the triplet hotspots, we move from the relative momentum, $\boldsymbol{p}$, to the relative coordinate of a Cooper pair, $\boldsymbol{\rho}$, via an inverse Wigner transformation, 
$
\mathcal{F}_{\sigma\sigma}(\boldsymbol{R},\boldsymbol\rho)
=
\int \frac{d^2p}{(2\pi\hbar)^2}
e^{\frac{i}{\hbar}\boldsymbol{p}\cdot\boldsymbol{\rho}}
\mathcal F_{\sigma\sigma}(\boldsymbol{R},\boldsymbol{p}).
$
To quantify the net triplet strength, we focus on small $\boldsymbol{\rho}$ and expand
$
\mathcal{F}_{\sigma\sigma}(\boldsymbol{R},\boldsymbol\rho)
\sim
\frac{i}{\hbar}
\boldsymbol \rho\cdot \int \frac{d^2p}{(2\pi\hbar)^2}\,
\boldsymbol{p}\,
\mathcal{F}_{\sigma\sigma}(\boldsymbol{R},\boldsymbol{p}),
$
where the integral expression corresponds to the 
short-range $p$-wave triplet amplitude. 
The lowest order contribution in the expansion of the exponential vanishes because the pairing correlator is odd in momentum, since 
$\alpha_{\boldsymbol p}\propto p_R$. 
We define the associated spin-resolved triplet intensity
\begin{equation}
I^{\sigma\sigma}_t(\boldsymbol R)
\equiv
\left|
\int \frac{d^2p}{(2\pi\hbar)^2}\,
\boldsymbol p\,
\mathcal F_{\sigma\sigma}(\boldsymbol R,\boldsymbol p)
\right|^2. 
\label{Eq12}
\end{equation}
Moreover, we introduce the total triplet intensity as 
$
I_t(\boldsymbol R)
\equiv
I^{\uparrow\uparrow}_t(\boldsymbol R)
+
I^{\downarrow\downarrow}_t(\boldsymbol R)
$
and the spin-selectivity as
$
S_t(\boldsymbol R)
\equiv
[I^{\uparrow\uparrow}_t(\boldsymbol R)
-
I^{\downarrow\downarrow}_t(\boldsymbol R)]
/
[I^{\uparrow\uparrow}_t(\boldsymbol R)
+
I^{\downarrow\downarrow}_t(\boldsymbol R)].
$

\begin{figure}[!t]
    \centering
    \includegraphics[width=1\linewidth]{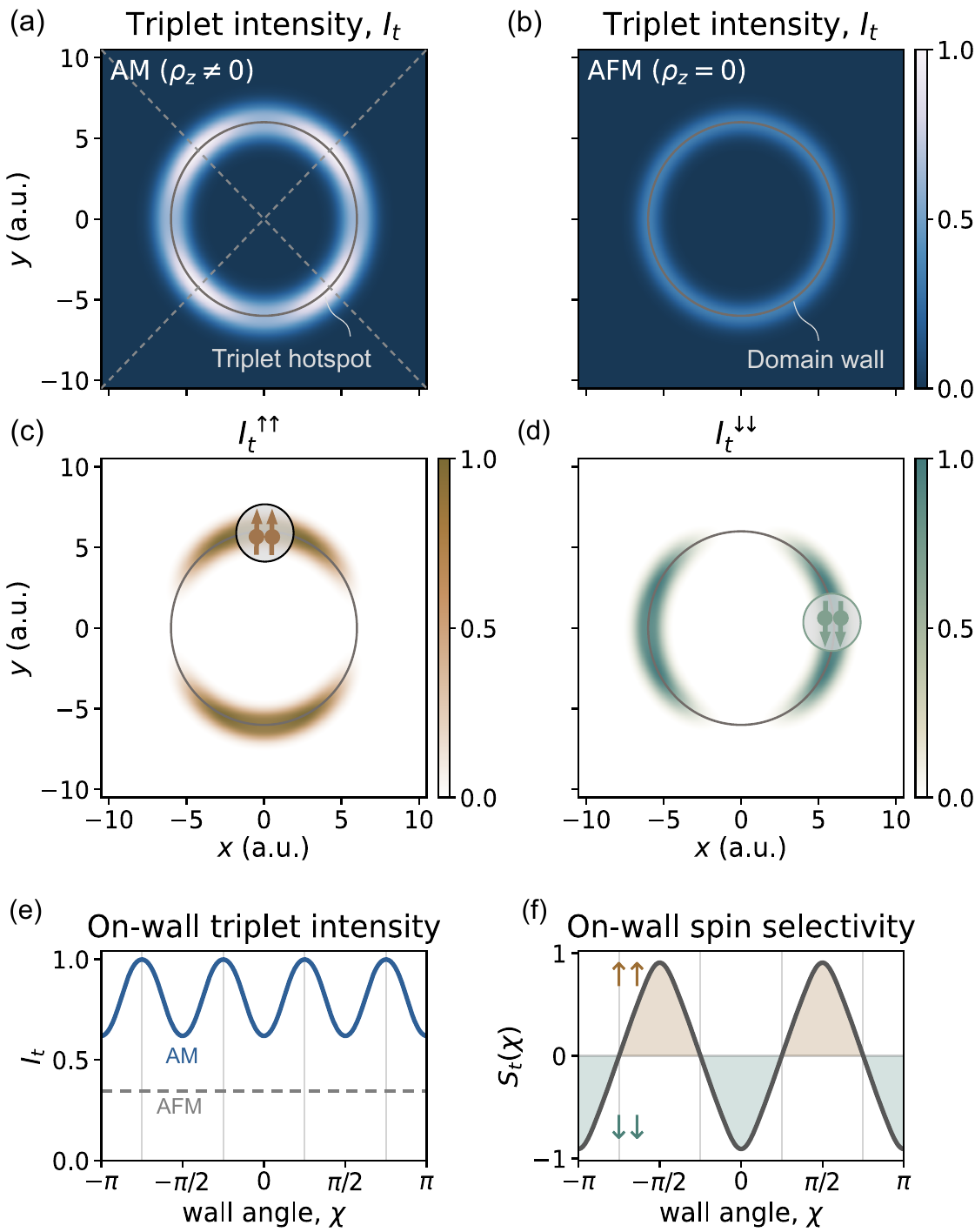}
\caption{
\textbf{Spin-polarized triplet hotspots at a proximitized radial altermagnetic domain wall.}
(a) Equal-spin triplet intensity, $I_t(\boldsymbol r)$, for a radial AM wall, $\varrho_z\neq0$. The wall generates triplets, while the $d$-wave AM splitting produces a fourfold hotspot pattern.
(b) Non-AM limit ($\varrho_z=0$). Triplets remain localized at the wall, but the angular modulation is absent.
(c,d) Spin-resolved intensities: $I_t^{\uparrow\uparrow}$ peaks near $\chi=\pm\pi/2$, whereas $I_t^{\downarrow\downarrow}$ peaks near $\chi=0,\pi$.
(e) On-wall cut of $I_t$, comparing AM and non-AM cases.
(f) On-wall spin selectivity $S_t(\chi)$, with positive (negative) values denoting $\uparrow\uparrow$ ($\downarrow\downarrow$) dominance.
}
    \label{fig:3}
\end{figure}

We have evaluated the triplet intensities and the spin-selectivity numerically. 
Our results are shown in Fig.~\ref{fig:3}. 
In the altermagnetic case, Fig.~\ref{fig:3}(a), the total triplet intensity is localized along the wall but is not uniform.
Instead, it forms a fourfold ``hotspot'' pattern. 
Notably, this angular modulation disappears in the antiferromagnetic limit, $\varrho_z=0$, shown in Fig.~\ref{fig:3}(b). 
This comparison thus shows that the emergent spin-orbit coupling generates equal spin-triplet 
correlations around the wall, while the altermagnetic anisotropy redistributes the equal spin-triplet 
correlations in a nonuniform way. 

We can also understand the formation of these triplet hotspots 
from the individual intensities of $\uparrow\uparrow$ and $\downarrow\downarrow$ pairs,
as shown in Fig.~\ref{fig:3}(c). 
The $\uparrow\uparrow$ intensity is largest near $\chi=\pm\pi/2$, 
whereas the $\downarrow\downarrow$ intensity is largest near $\chi=0,\pi$. 
The total intensity is enhanced where the spin-resolved intensities overlap, 
producing the total spin triplet hotspots of Fig.~\ref{fig:3}(e). 
Interestingly, at the same locations, 
the relative spin polarization is reduced because both spin species contribute.
This is reflected in the spin selectivity, Fig.~\ref{fig:3}(f), 
whose extrema occur along the principal axes where one equal-spin component dominates.

\section{Supercurrent-induced torques}
So far, we have seen that an altermagnetic texture can be used to induce spatially varying spin-triplet correlations. 
We now consider a converse question:
how does a supercurrent let an $s$-wave superconductor act back on the texture? 
This question is important because it asks how an altermagnetic texture can, in principle, be manipulated by a supercurrent.

To address this question, we stick to our example of the planar radial domain wall,
but assume that the applied supercurrent leads to a finite Cooper-pair momentum $\hbar\vec{Q}\neq 0$, i.e., set $\Delta(\boldsymbol{R}) \rightarrow e^{i\vec{Q}\cdot \vec{R}}\Delta(\boldsymbol{R})$. 
Within the semiclassical approximation, the BdG Hamiltonian (with appropriately re-defined Nambu spinors) is 
\begin{align}
&\mathcal{H}_{\text{BdG}}(\boldsymbol{R},\boldsymbol{p};\boldsymbol{Q})\\
&=
\begin{pmatrix}
h_{\text{proj}}(\boldsymbol{R},\boldsymbol{p}+\hbar\boldsymbol{Q}/2)
&
\Delta(\boldsymbol{R})(-i\sigma_y)
\\
-\Delta(\boldsymbol{R})(-i\sigma_y)
&
-[h_{\text{proj}}(\boldsymbol{R},-\boldsymbol{p}+\hbar\boldsymbol{Q}/2)]^T
\end{pmatrix}\nonumber,
\end{align}
where we chose a gauge with real-valued $\Delta(\boldsymbol{R})$. For brevity, we will now omit the $(\boldsymbol{R},\boldsymbol{p})$ arguments in the following. 

The supercurrent-induced torque follows from the supercurrent-induced change in the free energy density upon varying the angle $\phi(\vec{R})$ defining the local orientation $\boldsymbol n(\boldsymbol R)
=
(\cos\phi(\boldsymbol R),\sin\phi(\boldsymbol R),0)$ of the texture.  
Up to a texture-independent additional term, which does not affect the torque, the free energy density is given by, 
\begin{equation}
\Omega[\phi,\boldsymbol Q]
=
-\frac{T}{2}
\sum_{\omega_n}
\text{Tr}
\ln
\left[
i\omega_n
-
\mathcal H_{\text{BdG}}(\boldsymbol{Q})
\right].
\end{equation}
Here, the trace is over Nambu, spin, and phase space degrees of freedom. 
For small Cooper-pair momenta, we can expand the change in the free energy density due to the supercurrent, $\Omega_Q\equiv \Omega[\phi,\boldsymbol{Q}]-\Omega[\phi,\boldsymbol{0}]$. Up to second order in the Cooper-pair momentum, we have, 
\begin{align}
\Omega_Q
&\approx
\int d^2 R \,
[
Q J_Q(\boldsymbol{R})
+
\frac{Q^2}{2} D_Q(\boldsymbol{R})
]. 
\label{Eq15}
\end{align}
Here, a non-zero linear term, $J_Q(\boldsymbol{R})=\hat{Q}_i J^i(\boldsymbol{R})$, with the unit-vector components $\hat{Q}_i = \vec{Q}_i/Q$, would stabilize $Q \neq 0$ at equilibrium. 
The quadratic term,
$
D_Q(\boldsymbol{R})
=
\hat{Q}_i \hat{Q}_j D^{ij}(\boldsymbol{R})
$,
describes the superfluid stiffness along the supercurrent direction. 
For the radial domain wall, the linear term vanishes. 
To see this, we note that the BdG Hamiltonian obeys
$
\text{diag}(1,-1,-1,1)
\mathcal H_{\text{BdG}}(\boldsymbol{p},\boldsymbol{Q})
\text{diag}(1,-1,-1,1)
=
\mathcal H_{\text{BdG}}(-\boldsymbol{p},-\boldsymbol{Q})
$, which corresponds to spinful $\pi$ rotation along the $z$ axis.
After the momentum integration, the free energy is therefore even under 
$\boldsymbol{Q}\to-\boldsymbol{Q}$ and, therefore, cannot carry a finite supercurrent in equilibrium. 
The leading supercurrent-induced change to the free energy density is given by the superfluid stiffness, $D_Q(\boldsymbol{R})$. 

\begin{figure}[!t]
    \centering
    \includegraphics[width=1\linewidth]{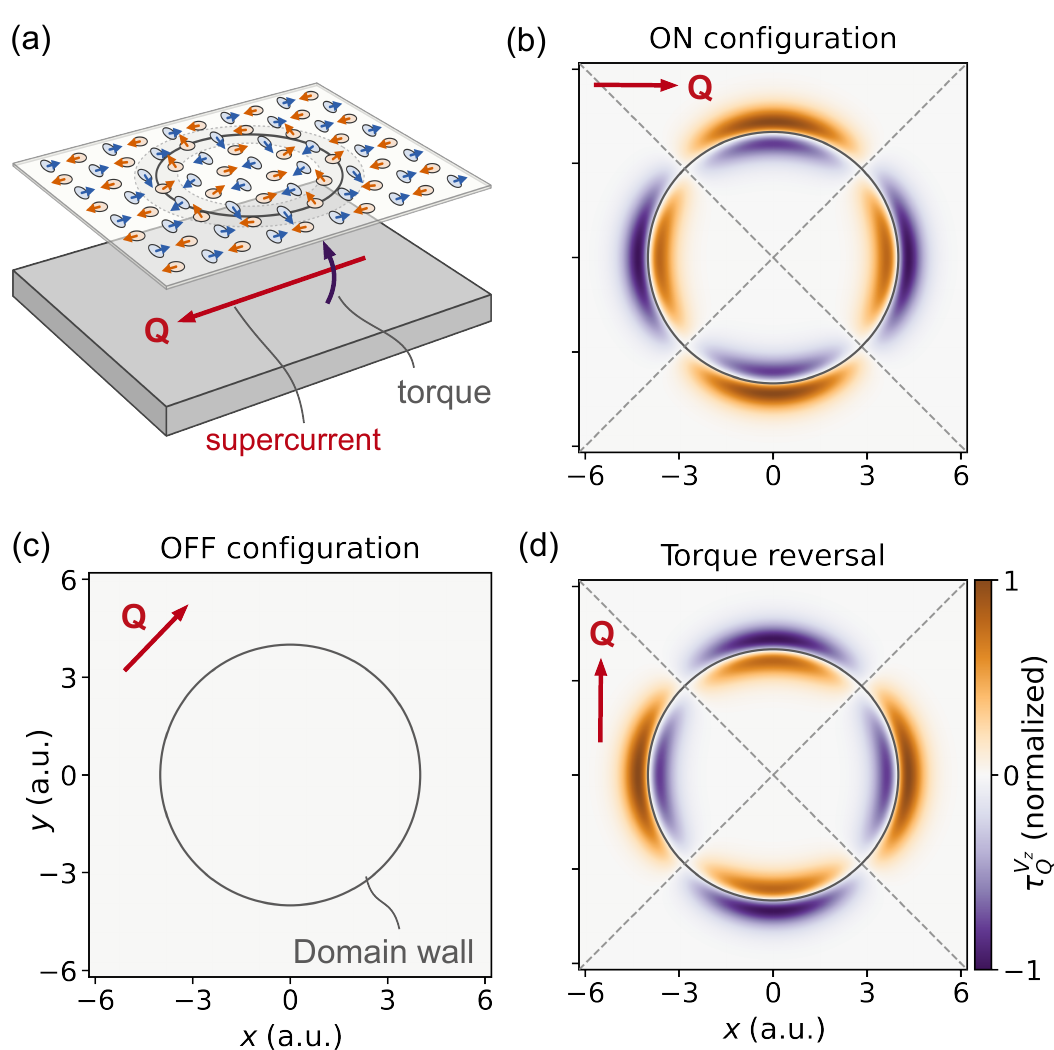}
\caption{
\textbf{Supercurrent-induced torque hotspots at a radial altermagnetic domain wall.}
(a) A supercurrent produces an out-of-plane torque on the in-plane Néel vector of the altermagnetic texture.
(b) Altermagnetic torque contribution, $\tau_Q^{V_z}$, for a current along the $x$ direction. The torque is localized near the domain wall and has the quadrupolar pattern set by the $d$-wave order parameter.
(c) For a diagonal current, $\chi_Q=\pi/4$, the altermagnetic torque vanishes.
(d) Rotating the current from $x$ to $y$ reverses the torque pattern. 
All torques are normalized by their maximum. 
}
    \label{fig:4}
\end{figure}

For a planar texture, the resulting torque has only an out-of-plane component
and is obtained by varying the change in the free energy density with respect to the texture angle, 
\begin{equation}
\tau_Q(\boldsymbol R)
=
-
\frac{\delta\Omega_Q}{\delta\phi(\boldsymbol R)}.
\label{Eq16}
\end{equation}

To evaluate this torque, we need the dependence of $D_Q$ on the texture-induced fields. 
For the planar radial wall, 
$
V_0=(\hbar^2/4)\delta^{ij}\partial_i\phi\,\partial_j\phi
$
,
$
V_z=(\hbar^2/4)\eta^{ij}\partial_i\phi\,\partial_j\phi
$,
and
$
\alpha_i=\hbar\varrho_3\partial_i\phi
$.
Assuming a slowly-varying texture, we expand the superfluid stiffness
up to the lowest non-vanishing order in these fields, 
\begin{equation}
D_Q
\approx
\bar{D}_Q
+
\varrho_0\bar C_{0,Q}V_0
+
\varrho_z\bar C_{z,Q}V_z
+
\frac{1}{2}\bar A_Q^{mn}\alpha_m\alpha_n.
\label{Eq17}
\end{equation}
Here, $\bar{D}_Q$, $\bar C_{0,Q}$, $\bar C_{z,Q}$, and $\bar A_Q^{mn}$ are the Taylor coefficients of the expansion, which need to be evaluated in the uniform limit, $V_0=V_z=\alpha_i=0$. 

Three remarks about this expansion are in order: First, 
the leading order contribution from the emergent spin-orbit coupling is quadratic. 
To see this, we note from Eq.~\eqref{Eq8} that sending 
$\boldsymbol{\alpha}\to-\boldsymbol{\alpha}$ is equivalent
to a $\pi$-rotation in spin space around the $z$-axis, implying that the spectrum must be invariant under
this transformation. Hence, the superfluid stiffness cannot contain terms that are odd in $\boldsymbol{\alpha}$.
Second, the expansion in Eq.~\eqref{Eq17} contains only terms up to second order in the 
gradients of $\phi$, while $\mathcal{O}[(\partial_i\phi)^4]$ terms are neglected. 
Third, the term in the expansion that is unique to the altermagnet is $\propto V_z$.

We now insert the expansion of Eq.~\eqref{Eq17} into the free energy density
and compute the torque by performing the variational derivative.
We find three contributions, one for each of the texture-induced fields, $\tau_Q
=
\tau_Q^{V_0}
+
\tau_Q^{V_z}
+
\tau_Q^{\text{SOC}}$. 
Specifically,
\begin{equation}
\begin{split}
\tau_Q^{V_0}
&=
\frac{Q^2}{2}
\partial_j
\left[
\frac{\hbar^2 \varrho_0}{2}
\bar{C}_{0,Q}
\delta^{ij}
\partial_i \phi
\right],
\\
\tau_Q^{V_z}
&=
\frac{Q^2}{2}
\partial_j
\left[
\frac{\hbar^2 \varrho_z}{2}
\bar{C}_{z,Q}
\eta^{ij}
\partial_i \phi
\right],
\\
\tau_Q^{\text{SOC}}
&=
\frac{Q^2}{2}
\partial_j
\left[
\hbar^2 \varrho_3^2
\bar{A}_Q^{jn}
\partial_n \phi
\right].
\end{split}    
\end{equation}
We will now focus on $\tau_Q^{V_z}$, since it
is unique to the altermagnet. 

For the radial wall, we can evaluate the altermagnetic contribution explicitly. 
We set
$\hat{\boldsymbol{Q}}=(\cos\chi_Q,\sin\chi_Q)$
and
$\boldsymbol{R}=R(\cos\chi,\sin\chi)$. 
The altermagnetic torque is then given by, 
\begin{equation}
\tau_Q^{V_z}
=
\frac{\hbar^2\varrho_zQ^2}{4}
\bar C_z
\cos 2\chi_Q
\cos 2\chi
\left[
\phi''(R)-\frac{\phi'(R)}{R}
\right].
\label{Eq19}
\end{equation}
Here, the coefficient $\bar C_z$ can be calculated from the spectrum of the proximitized uniform $d$-wave altermagnet. The BdG eigenvalues of the latter are 
given by 
$
\bar{E}_{s\tau}
=
s\varrho_z(p_x^2-p_y^2)
+
\tau[(\varrho_0p^2-\mu)^2+\Delta^2]^{1/2}
$
with $s,\tau=\pm$. We then find,
\begin{equation}
\bar C_z
=
\frac{\eta_{ij}}{4}
\int
\frac{d^2p}{(2\pi\hbar)^2}
\sum_{s,\tau}
s
[
f''(\bar{E}_{s\tau})
\bar v_{s\tau}^i
\bar v_{s\tau}^j
+
f'(\bar{E}_{s\tau})
\bar m_{s\tau}^{ij}
].
\label{Eq20}
\end{equation}
Here,
$\bar v_{s\tau}^i=\hbar\,\partial_{p_i}\bar{E}_{s\tau}$,
$\bar m_{s\tau}^{ij}=\hbar^2\,\partial_{p_i}\partial_{p_j}
\bar{E}_{s\tau}$, 
and $f(E)=1/(e^{\beta E}+1)$
is the Fermi function. 

Eq.~\eqref{Eq19} and Eq.~\eqref{Eq20} are the main results of this section. Several points about them are noteworthy: 

First, the supercurrent produces a quadrupolar altermagnetic torque around the domain wall, due to the $\cos2\chi$ factor that originates from the
$d$-wave order parameter. 
In addition, the factor $\cos2\chi_Q$ shows that the torque is controllable by the current direction.  
Currents along $x$ and $y$ generate opposite torques, 
while a diagonal current, for example $\chi_Q=\pi/4$, 
switches the altermagnetic torque off. 
The current direction thus controls both the sign and the magnitude of the altermagnetic torque.

Second, the torque is localized at the domain wall. 
The radial factor, 
$\phi''(R)-\phi'(R)/R$
is nonzero only where the Néel vector rotates.
The supercurrent therefore produces ``torque hotspots'' near the domain wall.

Third, the altermagnetic torque is a quasiparticle response. 
At zero temperature, $f'(E)$ and $f''(E)$ have support only at zero quasiparticle energy. 
Thus, in a zero-temperature fully gapped superconducting state, $\bar C_z=0$ and the altermagnetic torque vanishes. 
It remains finite though if the superconductor exhibits a 
Bogoliubov Fermi surface with $\bar{E}_{s\tau}(\boldsymbol p)=0$. 
At finite temperature, thermally excited quasiparticles can produce a finite torque even in the fully gapped regime.

\section{Domain wall deformation}
Finally, an interesting question is how the altermagnetic torque,
$
\tau_Q^{V_z},
$
can deform a radial domain wall. 
We will address this question by showing that the altermagnetic torque deforms a circular planar domain into an ellipse with an orientation that is set by the direction of the supercurrent.

We begin by relating a deformation of the wall position to a deformation of the altermagnetic texture. 
In the absence of the torque, we describe the planar domain wall in terms of a texture angle, 
$
\phi_0(r)=(\pi/2)\tanh[(r-R_0)/w].
$
Upon application of the altermagnetic torque, we assume that 
the domain wall radius deforms in an angle-dependent way, 
$
R_0\to R(\chi)=R_0+\delta u(\chi)
$.
This deformation is, to linear order, equivalent to a change of the texture angle, 
$
\phi_0(r)\rightarrow \phi_0(r)+\delta\phi(r,\chi)
$
with 
$
\delta\phi(r,\chi)
=
-\phi_0'(r)\delta u(\chi).
$

To realize such a texture deformation, the altermagnetic torque has to perform mechanical work on the texture. This work is given by,
\begin{equation}
\delta W_Q
=
\int d^2r
\,
\tau^{V_z}_Q(r,\chi)
\delta\phi(r,\chi).
\label{Eq21}
\end{equation}
Inserting the expression for the altermagnetic torque from Eq.~\eqref{Eq19}, we see that this work has a quadrupolar form, 
$
\delta W_Q\propto
\int d\chi\,\cos2\chi\,\delta u(\chi),
$
up to an overall radial prefactor. 

To demonstrate this elliptical deformation, it is now useful to decompose the 
wall displacement into angular harmonics. We write
$
\delta u(\chi)
=
\delta u_0
+
\delta u_{1}\cos\chi
+
\delta u'_{1}\sin\chi
+
\delta u_{2}\cos2\chi
+
\delta u'_{2}\sin2\chi
+\cdots.
$
Here, $\delta u_0$ describes a uniform expansion of the wall. 
The two first harmonics,
$
\delta u_{1}\cos\chi
$
and
$
\delta u'_{1}\sin\chi
$,
describe shifts of the domain wall in the $x$- and $y$-directions. 
The second harmonics describe quadrupolar distortions of the wall. 
By the orthogonality of the angular harmonics, all terms vanish in $\delta W_Q$ except the term 
$
\propto
\delta u_2\cos2\chi
$.
This term will give rise to the elliptical deformation.

We will now evaluate the work for the non-zero deformation mode. 
We set 
$
\delta u(\chi)=\delta u_2\cos2\chi
$
and perform the radial and angular integrals in Eq.~\eqref{Eq21}. We find that,
\begin{equation}
\begin{split}
\delta W_Q
&=
F_Q\,\delta u_2,
\\
F_Q
&=
\frac{\pi^3}{8w^2}
\hbar^2\varrho_z Q^2
\bar C_z
\cos2\chi_Q .
\end{split}
\end{equation}
Here, $F_Q$ is a generalized force acting on the displacement $\delta u_2$. From the $\cos2\chi_Q$ dependence, we see that this generalized force is controlled by the direction of the applied supercurrent. 

We can now determine the equilibrium shape of the domain wall by balancing the generalized force 
$F_Q$ with a phenomenological restoring force from the domain wall. We write this
restoring force as 
$
F_{\text{rest}}=-k\delta u_2
$,
where $k$ is a phenomenological stiffness parameter. 
Force balancing then requires
$
F_Q+F_{\text{rest}}=0.
$
This constraint gives the following equilibrium shape of the domain wall,
\begin{equation}
\begin{split}
R(\chi)
&=
R_0+\delta u_2\cos2\chi,
\\
\delta u_2
&=
\frac{\pi^3}{8w^2 k}
\hbar^2\varrho_z Q^2
\bar C_z
\cos2\chi_Q .
\end{split}
\end{equation}
Hence, for sufficiently small $\delta u_2$, the resulting equilibrium shape of the domain wall is elliptical. The elongation of the ellipse is set by the sign of $\delta u_2$ or, equivalently, by the supercurrent direction, $\cos2\chi_Q$.
For $\delta u_2 > 0$, the ellipse will be elongated along $x$, while for $\delta u_2  < 0$ it will by elongated along $y$. Importantly, the supercurrent does not allow for a continuous rotation of the ellipse. Instead, the possible elongation directions are fixed by the altermagnetic anisotropy.

\begin{figure}[!t]
    \centering
    \includegraphics[width=1\linewidth]{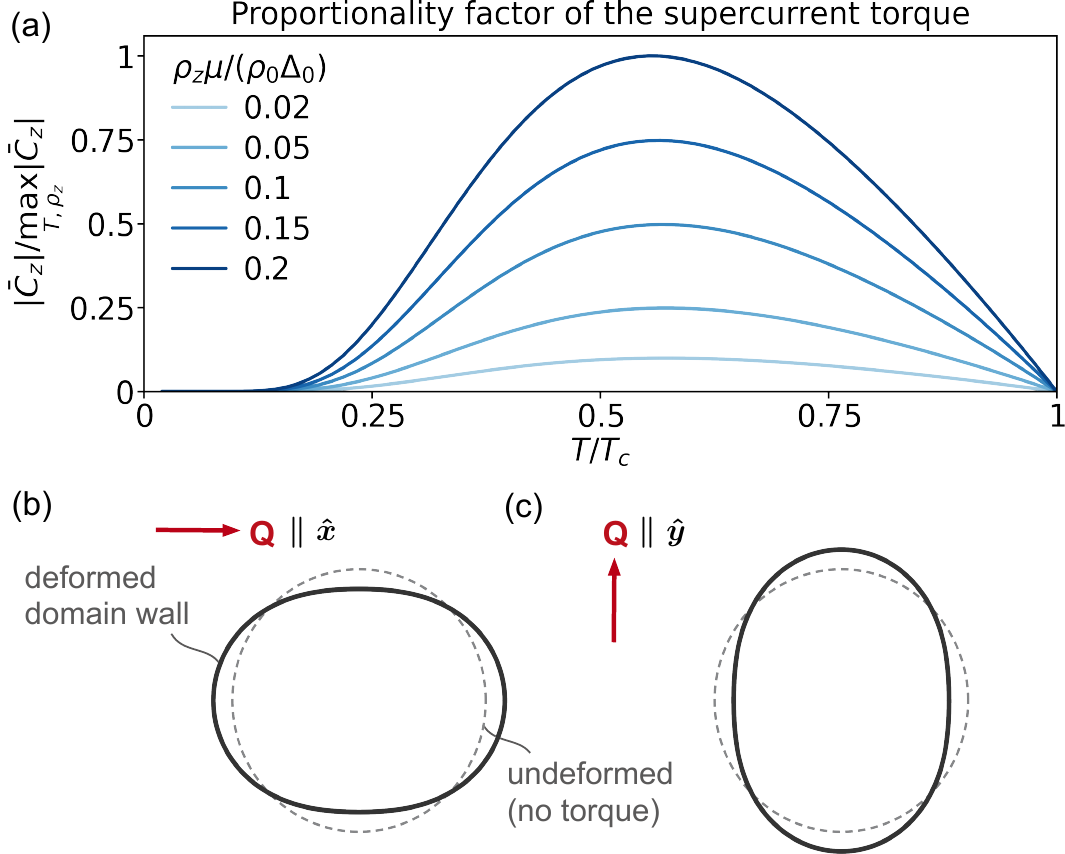}
\caption{
\textbf{Supercurrent torque coefficient and current-controlled domain-wall deformation.}
(a) Temperature dependence of the altermagnetic torque coefficient, $\bar{C}_z$, normalized to its maximum in the plotted range.
Curves show different altermagnetic splittings, $\varrho_z\mu/(\varrho_0\Delta_0)$.
The torque is quasiparticle-mediated.
It vanishes at $T=0$ for a fully gapped spectrum and is suppressed again as the gap closes near $T_c$.
(b,c) The torque deforms a radial domain wall into an ellipse.
A supercurrent with Cooper-pair momentum $\boldsymbol Q\parallel\hat{\boldsymbol x}$ deforms the radial domain wall along one principal axis of the $d_{x^2-y^2}$ altermagnetic form factor. A supercurrent with $\boldsymbol Q\parallel\hat{\boldsymbol y}$ deforms the wall along the other principal axis. Plots are schematic.
}
    \label{fig:5}
\end{figure}

\section{Discussion and Conclusion}
In this work, we have shown that domain wall textures or other forms of order parameter inhomogeneities, which are expected to typically be present in altermagnetic samples, can play an important role for superconductivity:

First, while the time-reversal-odd spin-splitting of the Fermi surfaces in a homogeneous altermagnet naturally suppresses the formation of a superconducting gap from proximity-induced singlet pairing, domain-wall boundaries locally induce triplet components due to the emergent spin-orbit coupling; as we have shown, this gaps out the BdG spectra in the vicinity of the domain wall and the spatial asymmetry of the altermagnet leads to remaining nodal points in momentum space or fully gapped spectra, depending on the position on the domain wall. As such, while the proximity effect from a superconductor into an altermagnet might not lead to significant global low-energy pairing correlations across the whole altermagnet, scanning probes, in particular scanning tunneling microscopy, could be used to pick up the local gap formation around domain walls. Consequently, one would expect a transition from gapless spectra in homogeneous regions of the altermagnet to angular-dependent transitions from V-shaped to U-shaped tunneling spectra around domain walls.

Second, our findings also have implications for intrinsic, i.e., interaction-induced, superconductivity in the altermagnet itself: in the scenario with dominant interactions in the spin-singlet Cooper-pair channel, a homogeneous altermagnetic order parameter can suppress pairing; however, it might still nucleate around domain wall boundaries based on the mechanism we discuss here, which should again be clearly visible in scanning probe experiments. Although in a very different regime, we note the similarity to the prediction~\cite{PhysRevB.111.L100502,2025arXiv251019943S} that local, non-magnetic defects can enhance pairing in altermagnets. 

Third, it is instructive to provide a material estimate for the importance of the proposed 
spin triplet pairing. We focus on Mn$_{5}$Si$_{3}$, as a candidate $d$-wave altermagnet~\cite{reichlova}. 
The ratio between the spin triplet pairing, $\Delta_t$, and the proximity-induced gap, $\Delta$, 
is approximately 
$\Delta_{t}/\Delta
\sim
\alpha/[\sqrt{\alpha^{2}+
(\Delta_{\text{AM}}/2)^{2}}]
$,
where $\alpha$ is an estimate for the emergent spin-orbit coupling and 
$\Delta_{\text{AM}}$ for the altermagnetic band splitting.
At the domain wall radius,
we estimate 
$\alpha\sim\pi\hbar v_{3}/4w$ 
with $v_{3}= 2\varrho_{3}p_{F}$  being a velocity scale 
and 
$p_{F}$ 
being a Fermi momentum scale.
We take 
$v_{F}\sim 2\times 10^{6}~\mathrm{m\,s^{-1}}$~\cite{nernst}
and assume $v_{3}\sim 0.5v_{F}$.
For a wall width $w\sim50~$nm, we find $\alpha\sim10.3$~meV.
For $\Delta_{\text{AM}}\sim 100$~meV~\cite{dft}, this gives $\Delta_t/\Delta\sim0.2$. 
Assuming a proximity-induced gap of $\Delta\sim0.5$~meV, we find a triplet pairing contribution
of $\Delta_t\sim0.1$~meV. 

Finally, in addition to the formation of spin-triplet superconductivity, we have also shown 
that the application of a supercurrent can lead to torques on the textures. In particular, 
the resulting deformations of the domain wall structure depend on the structure of the altermagnetic order parameter. These results point to new ways of controlling altermagnetic textures with supercurrents.

\begin{acknowledgments}
M.S.S. thanks S.~Banerjee and U.~Seifert for insightful discussions and acknowledges funding by the European Union (ERC-2021-STG, Project 101040651— SuperCorr). Views and opinions expressed are however those of the authors only and do not necessarily reflect those of the European Union or the European Research Council Executive Agency. Neither the European Union nor the granting authority can be held responsible for them. C.S. acknowledges support from the Louisiana Board of Regents. This work was performed in part at the Aspen Center for Physics, which is supported by National Science Foundation grant PHY-2210452.
\end{acknowledgments}

\onecolumngrid

\newpage

\clearpage
\setcounter{page}{1}
\renewcommand{\thepage}{\arabic{page}}

\setcounter{section}{0}
\renewcommand{\thesection}{S\arabic{section}}
\renewcommand{\thesubsection}{S\arabic{section}.\arabic{subsection}}
\renewcommand{\thesubsubsection}{S\arabic{section}.\arabic{subsection}.\arabic{subsubsection}}

\setcounter{figure}{0}
\setcounter{table}{0}
\renewcommand{\thefigure}{S\arabic{figure}}
\renewcommand{\thetable}{S\arabic{table}}

\setcounter{equation}{0}
\renewcommand{\theequation}{S\arabic{equation}}

\begin{center}
\large{\bf Supplemental Material \\}
\end{center}
\begin{center}Yasir Dar$^{1}$, Mathias S. Scheurer$^{2}$, and Constantin Schrade$^{1}$
\\
{\it $^{1}$Hearne Institute of Theoretical Physics, Department of Physics \& Astronomy, Louisiana State University, Baton Rouge LA 70803, USA}
\\
{\it $^{2}$Institute for Theoretical Physics III, University of Stuttgart, 70550 Stuttgart, Germany}
\end{center}

\section{Effective Bogoliubov-de Gennes Hamiltonian}
In this first section, we provide more details on the derivation of the  effective low-energy Bogoliubov-de Gennes (BdG) Hamiltonian for an altermagnetic texture proximitized by an $s$-wave superconductor.

\subsection{Normal-state Hamiltonian}
For convenience, we begin by redisplaying the normal state Hamiltonian for our altermagnetic spin texture,
\begin{equation}
\hat{h}(\boldsymbol{r},\hat{\boldsymbol{p}})
=
\varepsilon_{0,\hat{\boldsymbol{p}}}
+t_{x,\hat{\boldsymbol{p}}}\tau_x
+t_{z,\hat{\boldsymbol{p}}}\tau_z
+J\,\tau_z\,\boldsymbol{n}(\boldsymbol{r})\cdot\boldsymbol{s}
-\mu,
\end{equation}
Here, the Pauli matrices, $\tau_{x,y,z}$, act in sublattice space and the Pauli matrices $s_{x,y,z}$ act in spin space. Moreover, $\hat{\boldsymbol p}=-i\hbar\nabla$ is the momentum operator, $\mu$ is the chemical potential, and $\boldsymbol n(\boldsymbol{r})$ is the spatially-varying Néel vector. We will assume that the exchange coupling, $J>0$, is the dominant energy scale. 
For the kinetic terms, we will focus on a $d$-wave altermagnetic spin texture,
$
\varepsilon_{0,\hat{\boldsymbol{p}}} = \varrho_0 (\hat{p}_x^2 + \hat{p}_y^2)
$
,
$
t_{z,\hat{\boldsymbol{p}}} = \varrho_z (\hat{p}_x^2 - \hat{p}_y^2)
$,
and
$
t_{x,\hat{\boldsymbol{p}}} = \varrho_x + \varrho_3 (\hat{p}_x^2 + \hat{p}_y^2)
$.

\subsection{Rotating frame}
As a first step, we move to a rotation frame where the local spin quantization axis aligns with 
$\boldsymbol n(\boldsymbol r)$. 
\\
\\
The unitary transformation to the rotating frame is defined through the requirement,
$
U^\dagger(\boldsymbol{r}) 
\left[ \boldsymbol{n}(\boldsymbol{r}) \cdot \boldsymbol{s} \right] 
U(\boldsymbol{r})
= s_z
$.
We can write this transformation explicitly by parametrizing the Néel vector as, 
\begin{equation}
\boldsymbol{n}(\boldsymbol{r})
=
\left( 
\sin\theta \cos\phi,\ \sin\theta \sin\phi,\ \cos\theta \right)
\quad \text{so that} \quad
U(\boldsymbol{r})
\equiv
\exp\left[
-\frac{i}{2}\theta(\boldsymbol{r})
\left( 
-\sin\phi(\boldsymbol{r})
s_x 
+
\cos\phi(\boldsymbol{r})
s_y 
\right)
\right].
\end{equation}
Because $U(\boldsymbol{r})$ is spatially varying, the momenta in the texture Hamiltonian will be shifted by a gauge potential, 
\begin{equation}
A_i(\boldsymbol{r})
=
i \hbar
U^\dagger(\boldsymbol{r}) \partial_i U(\boldsymbol{r})
\quad \text{with} \quad
\hat{\pi}_i
\equiv
U^\dagger(\boldsymbol{r}) \hat{p}_i U(\boldsymbol{r})
=
\hat{p}_i - A_i(\boldsymbol{r}) .
\end{equation}
The resulting normal-state Hamiltonian in the rotating frame is thus given by,
\begin{equation}
\hat{h}_\text{rot}(\boldsymbol{r}, \hat{\boldsymbol{\pi}})
\equiv
U^\dagger(\boldsymbol{r}) \, \hat{h}(\boldsymbol{r}, \hat{\boldsymbol{p}}) \, U(\boldsymbol{r})
=
\varepsilon_{0,\hat{\boldsymbol{\pi}}}
+ t_{x,\hat{\boldsymbol{\pi}}}\,\tau_x
+ t_{z,\hat{\boldsymbol{\pi}}}\,\tau_z
+ J\,\tau_z s_z
- \mu,
\end{equation}
as already discussed in the main text. 
\\
\\
We now add a technical consideration. We decompose 
the gauge field into components that are either parallel or perpendicular to the spin quantization axis,
$
A_i(\boldsymbol{r})
=
A_i^{\parallel}(\boldsymbol{r}) + A_i^{\perp}(\boldsymbol{r})
$
with
$
A_i^{\parallel} \propto s_z,
$
and
$
A_i^{\perp} \propto s_x, s_y .
$
Such a decomposition allows us to introduce the longitudinally shifted momentum, 
$
\hat{\Pi}_i
\equiv
\hat{p}_i - A_i^{\parallel}(\boldsymbol{r})
$
so that
$
\hat{\pi}_i
=
\hat{\Pi}_i - A_i^{\perp}(\boldsymbol{r}) 
$.
Using the texture parametrization given above, we can write the longitudinal and transversal components as, 
\begin{equation}
A_i^{\parallel}(\boldsymbol{r})
=
- \hbar\, (\partial_i \phi)\, \sin^2\left(\frac{\theta}{2}\right) s_z,
\quad \text{and} \quad
A_i^{\perp}(\boldsymbol{r})
=
\frac{\hbar}{2}
\left[
(\partial_i \theta)\, \boldsymbol{e}_{\phi}
- \sin\theta\, (\partial_i \phi)\, \boldsymbol{e}_{r}
\right]
\cdot (s_x, s_y) .
\end{equation}
with the unit vectors $\boldsymbol{e}_r=(\cos\phi,\sin\phi)^T$ and $\boldsymbol{e}_\phi=(-\sin\phi,\cos\phi)^T$. We can also write the transversal component as, 
\begin{equation}
\begin{split}
A_i^\perp(\boldsymbol r)  =\hbar\sum_{a=x,y} e_{ia}(\boldsymbol r)\,s_a
\quad \text{with} \quad
e_{ix}(\boldsymbol r) &=
-
\frac{1}{2}\left[(\partial_i\theta)\sin\phi
+\sin\theta(\partial_i\phi)\cos\phi\right],
\\
e_{iy}(\boldsymbol r)
&=
\frac{1}{2}\left[(\partial_i\theta)\cos\phi
-\sin\theta(\partial_i\phi)\sin\phi\right].
\end{split}    
\end{equation}

\subsection{Bogoliubov-de Gennes Hamiltonian}
As a second step, we introduce the superconducting pairing and the BdG Hamiltonian. 
We define $\hat{\psi}(\boldsymbol r)=(\hat{c}_{+,\uparrow},\,\hat{c}_{+,\downarrow},\,\hat{c}_{-,\uparrow},\,\hat{c}_{-,\downarrow})^T$ as the electron spinor in spin- and sublattice-space, and $\hat{\Psi}(\boldsymbol r)=(\hat{\psi}(\boldsymbol r), \hat{\psi}^\dagger(\boldsymbol r)^T)^T$ as the Nambu spinor. In this basis, the BdG Hamiltonian is given by,
\begin{equation}
\hat{\mathcal{H}}_{\text{BdG}}(\boldsymbol{r})
=
\begin{pmatrix}
\hat{h}(\boldsymbol{r}, \hat{\boldsymbol{p}}) & \hat{\Delta}(\boldsymbol{r}) \\
\hat{\Delta}^\dagger(\boldsymbol{r}) & -\hat{h}^{T}(\boldsymbol{r}, -\hat{\boldsymbol{p}})
\end{pmatrix}.
\end{equation}
We will choose the pairing matrix to be of the form, 
\begin{equation}
\hat\Delta(\boldsymbol{r})=\Delta(\boldsymbol{r})\,\tau_x(i s_y).
\end{equation}
As discussed in the main text, this pairing matrix corresponds to inter-sublattice spin-singlet pairing and will be non-zero upon projection onto the low-energy subspace of the normal-state Hamiltonian. To introduce this low-energy subspace, we define the operators $S\equiv \tau_z s_z$ and the corresponding projectors $P_\pm \equiv \frac{1}{2}(1\pm S)$. We associate $S=-1$ with the low-energy subspace with projector $P_{-}$,
and $S=+1$ with the high-energy subspace with projector $P_{+}$.

\subsection{Bogoliubov-de Gennes Hamiltonian in the rotating frame}
As a third step, we transform the BdG Hamiltonian into the rotating frame. We define,
\begin{equation}
\mathcal{U}(\boldsymbol{r})=
\begin{pmatrix}
U(\boldsymbol{r}) & 0\\
0 & U^{*}(\boldsymbol{r})
\end{pmatrix}
\quad \text{so that} \quad
\hat{\mathcal{H}}_{\text{BdG,rot}}
\equiv
\mathcal{U}^{\dagger}\hat{\mathcal{H}}_{\text{BdG}}\mathcal{U}.
\end{equation}
We note that $U^{\dagger}(\boldsymbol{r})(i s_{y})U^{*}(\boldsymbol{r})= i s_{y}$. As a result, the pairing term is unchanged upon moving to the rotating frame. The BdG Hamiltonian in the rotating frame reads,
\begin{equation}
\hat{\mathcal{H}}_{\text{BdG,rot}}(\boldsymbol{r})
=
\begin{pmatrix}
\hat{h}_{\text{rot}}(\boldsymbol{r},\hat{\boldsymbol{\pi}}) &
\Delta(\boldsymbol{r})\tau_{x}(i s_{y})
\\
-\Delta^{*}(\boldsymbol{r})\tau_{x}(i s_{y}) &
-[\hat{h}_{\text{rot}}(\boldsymbol{r},-\hat{\boldsymbol{\pi}})]^{T}
\end{pmatrix}.
\end{equation}

\subsection{Low-energy effective Bogoliubov-de Gennes Hamiltonian}
As a fourth step, we project the BdG Hamiltonian onto the $S=-1$ subspace.
\\
\\
To formulate the projection, it is initially useful to define Pauli matrices, $\sigma_{x,y,z}$, that act in the low-energy $S=-1$ subspace spanned by $|1\rangle$ and $|2\rangle$. 
These Pauli matrices can be defined as,
$
\sigma_{0} \equiv P_{-} \tau_{0} s_{0} P_{-}
$
,
$
\sigma_{x} \equiv P_{-} \tau_{x} s_{x} P_{-}
$,
$
\sigma_{y} \equiv - P_{-} \tau_{x} s_{y} P_{-}
$
,
and
$
\sigma_{z} \equiv P_{-} \tau_{z} s_{0} P_{-} 
$.
\\
\\
Next, we define the projector onto the low-energy subspace, 
\begin{equation}
\mathcal{P}_{-} \equiv
\begin{pmatrix}
P_{-} & 0\\
0 & P_{-}
\end{pmatrix}.
\end{equation}
The low-energy effective BdG Hamiltonian is then given by, 
\begin{equation}
\hat{\mathcal{H}}_{\text{BdG,proj}}
\equiv
\mathcal{P}_{-}\,\hat{\mathcal{H}}_{\text{BdG,rot}}\,\mathcal{P}_{-}
=
\begin{pmatrix}
\hat{h}_{\text{proj}}(\boldsymbol{r},\hat{\boldsymbol{\Pi}})
&
\Delta(\boldsymbol{r})\,(-i\sigma_{y})
\\
-\Delta^{*}(\boldsymbol{r})\,(-i\sigma_{y})
&
-[\hat{h}_{\text{proj}}(\boldsymbol{r},-\hat{\boldsymbol{\Pi}})]^{T}
\end{pmatrix}.
\end{equation}
Here, the diagonal components involve the projected normal-state Hamiltonian, 
\begin{equation}
\hat{h}_{\text{proj}}(\boldsymbol{r},\hat{\boldsymbol{\Pi}})
=
\varepsilon_{0,\hat{\boldsymbol{\Pi}}}
+t_{z,\hat{\boldsymbol{\Pi}}}\,\sigma_{z}
+\varrho_{0} V_{0}(\boldsymbol{r})
+\varrho_{z} V_{z}(\boldsymbol{r})\,\sigma_{z}
-\mu
+\hat{h}^{(0)}_{\text{SOC}}(\boldsymbol{r}).
\end{equation}
with the emergent spin-orbit coupling, 
\begin{equation}
\begin{split}
\hat{h}_{\text{SOC}}(\boldsymbol{r})
&=
-\varrho_{3}\,P_{-}\,\tau_{x}\{\hat{p}_{i},A_{i}^{\perp}\}\,P_{-} 
\\
&=
-\hbar \varrho_{3}
\sum_{i=x,y}
\left(
\{\hat{p}_{i},e_{ix}(\boldsymbol{r})\}\sigma_{x}
-
\{\hat{p}_{i},e_{iy}(\boldsymbol{r})\}\sigma_{y}
\right).
\end{split}
\end{equation}
which arises from $\tau_{x}\hat{\boldsymbol{\pi}}^{2}
=
\tau_{x}\hat{\boldsymbol{\Pi}}^{2}
-\tau_{x}\{\hat{\Pi}_{i},A_{i}^{\perp}\}
+\tau_{x} A_{i}^{\perp} A_{i}^{\perp}$ and noting that $\{\hat{\Pi}_{i},A_{i}^{\perp}\}=\{\hat{p}_{i},A_{i}^{\perp}\}$.
\\
\\
The projected normal-state Hamiltonian also contains the quantum metric and the scalar potentials, 
\begin{equation}
g_{ij}(\boldsymbol{r})
\equiv
\frac{1}{4}\, \partial_{i} \boldsymbol{n}(\boldsymbol{r}) \cdot \partial_{j} \boldsymbol{n}(\boldsymbol{r}),
\quad
V_{0}(\boldsymbol{r})
\equiv
\hbar^{2} \delta^{ij} g_{ij}(\boldsymbol{r}),
\quad
V_{z}(\boldsymbol{r})
\equiv
\hbar^{2} \eta^{ij} g_{ij}(\boldsymbol{r}),
\quad
\eta = \text{diag}(1, -1) .
\end{equation}

\subsection{Radial domain wall and removal of the gauge potential}
As a fifth step, we focus on the example of a radial domain wall texture. For this texture, we will show that the longitudinal gauge field can be removed through a unitary transformation, so that the low-energy BdG Hamiltonian takes on a particularly simple form. 
\\
\\
We initially parametrize the radial domain wall texture as,
\begin{equation}
\boldsymbol{n}(\boldsymbol{r})=(\cos\phi(r),\sin\phi(r),0) 
\quad
\text{with}
\quad
\theta(r)=\frac{\pi}{2}\quad
\text{and}
\quad
\phi(r)=\frac{\pi}{2}\tanh\left(\frac{r-R_{0}}{w}\right),
\end{equation}
where we have introduced the parametrization $\boldsymbol{r}=(r\cos\chi,r\sin\chi)$ with 
$r=\sqrt{x^{2}+y^{2}}$. 
\\
\\
For this radial domain wall texture, the longitudinal gauge field and the shifted momenta are,
\begin{equation}
 A_{i}^{\parallel}(\boldsymbol{r})
=
\frac{\hbar}{2}(\partial_{i}\phi)\sigma_{z}
\quad
\text{and}
\quad
\hat{\Pi}_{i}
=
\hat{p}_{i}-\frac{\hbar}{2}(\partial_{i}\phi)\sigma_{z},
\end{equation}
where we have used that $P_{-} s_{z} P_{-}=-\sigma_{z}$.
\\
\\
An important simplification for the radial domain wall example is that the longitudinal gauge field can be removed from the kinetic terms via the unitary, 
\begin{equation}
W(\boldsymbol{r})
\equiv
e^{-\frac{i}{2}\phi(\boldsymbol{r})\sigma_{z}},   
\end{equation}
which satisfies the following properties,
\begin{equation}
W(\boldsymbol{r}) \hat{\Pi}_{i} W^{\dagger}(\boldsymbol{r})=\hat{p}_{i}
\quad
\text{and}
\quad
W\hat{h}^{(0)}_{\text{SOC}}W^{\dagger}
=
\frac{\hbar \varrho_{3}}{2}
\sum_{i=x,y}
\{\hat{p}_{i},\partial_{i}\phi(\boldsymbol{r})\}\sigma_{x}.
\end{equation}
To show these properties, we have used that 
$W^{\dagger} \sigma_{x} W
=
\cos\phi\;\sigma_{x}-\sin\phi\;\sigma_{y}$.
\\
\\
We apply this transformation to the Bogoliubov-de Gennes Hamiltonian via, 
\begin{equation}
\mathcal{W}(\boldsymbol{r})
=
\begin{pmatrix}
W(\boldsymbol{r}) & 0\\
0 & W^{*}(\boldsymbol{r})
\end{pmatrix}.
\end{equation}
The transformed Bogoliubov-de Gennes Hamiltonian is given by,
\begin{equation}
\hat{\mathcal{H}}_{\text{BdG,proj}}(\boldsymbol{r},\hat{\boldsymbol{p}})
\rightarrow
\mathcal{W}(\boldsymbol{r})
\hat{\mathcal{H}}_{\text{BdG,proj}}(\boldsymbol{r},\hat{\boldsymbol{\Pi}})
\mathcal{W}^{\dagger}(\boldsymbol{r})
=
\begin{pmatrix}
\hat{h}_{\text{proj}}(\boldsymbol{r},\hat{\boldsymbol{p}})
&
\Delta(\boldsymbol{r})(-i\sigma_{y})
\\
-\Delta^{*}(\boldsymbol{r})(-i\sigma_{y})
&
-\hat{h}_{\text{proj}}(\boldsymbol{r},-\hat{\boldsymbol{p}})^{T}
\end{pmatrix}
\end{equation}
with the transformed normal-state Hamiltonian, 
\begin{equation}
\hat{h}_{\text{proj}}(\boldsymbol{r},\hat{\boldsymbol{p}})
\rightarrow
\varepsilon_{0,\hat{\boldsymbol{p}}}
+t_{z,\hat{\boldsymbol{p}}}\sigma_{z}
+\varrho_{0}V_{0}(\boldsymbol{r})
+\varrho_{z}V_{z}(\boldsymbol{r})\sigma_{z}
-\mu
+
\frac{\hbar \varrho_{3}}{2}
\sum_{i=x,y}
\{\hat{p}_{i},\partial_{i}\phi(\boldsymbol{r})\}\sigma_{x}.  
\end{equation}

\newpage

\section{Anomalous Green's function}
In this section, we provide more details on the derivation of the anomalous Green's function for the radial altermagnetic texture discussed in the main text. 

\subsection{Coordinates and normal state Hamiltonian}
As a first step, we parameterize the center-of-mass coordinate and relative momentum, 
\begin{equation}
\boldsymbol{R}=R(\cos\chi,\sin\chi),
\quad
\text{and}
\quad
\boldsymbol{p}=p(\cos(\chi+\delta),\sin(\chi+\delta)).    
\end{equation}
With this choice, we have 
$p_{R}=\boldsymbol{p}\cdot\hat{\boldsymbol{R}}=p\cos\delta$
and
$p_{x}^{2}-p_{y}^{2}=p^{2}\cos2(\chi+\delta)$. 
The semiclassical normal state Hamiltonian then takes on the form, 
\begin{equation}
h_{\text{proj}}(\boldsymbol{R},\boldsymbol{p})
=
\xi(\boldsymbol{R},\boldsymbol{p})
+
\alpha_{\boldsymbol{p}}(\boldsymbol{R})\sigma_{x}
+
b_{z}(\boldsymbol{R},\boldsymbol{p})\sigma_{z} 
\end{equation}
with
\begin{equation}
\begin{split}
\xi(\boldsymbol{R},\boldsymbol{p})
&=
\varrho_{0} p^{2}
+
\frac{\hbar^{2}\varrho_{0}}{4}\phi'(R)^{2}
-\mu    
,
\hspace{3cm}
\alpha_{\boldsymbol{p}}(\boldsymbol{R})
=
\boldsymbol{p}\cdot\boldsymbol{\alpha}(\boldsymbol{R})
=
\hbar\varrho_{3}\phi'(R)p\cos\delta,
\\
b_{z}(\boldsymbol{R},\boldsymbol{p})
&=
\varrho_{z} p^{2}\cos2(\chi+\delta)
+
\frac{\hbar^{2}\varrho_{z}}{4}\phi'(R)^{2}\cos2\chi.
\end{split}    
\end{equation}

\subsection{Evaluating the anomalous Green's function}
As a second step, we obtain the anomalous Green's function from the off-diagonal components of the Nambu space Green's function as given in the main text, 
\begin{equation}
\mathcal{G}(\boldsymbol{R},\boldsymbol{p};i\omega_{n})
\approx
\big[
i\omega_{n}-\mathcal{H}_{\text{BdG}}(\boldsymbol{R},\boldsymbol{p})
\big]^{-1}
\quad
\text{with}
\quad
\mathcal{H}_{\text{BdG}}(\boldsymbol{R},\boldsymbol{p})
=
\begin{pmatrix}
h_{\text{proj}}(\boldsymbol{R},\boldsymbol{p})
&
\Delta(\boldsymbol{R})(-i\sigma_{y})
\\
-\Delta(\boldsymbol{R})(-i\sigma_{y})
&
-h_{\text{proj}}(\boldsymbol{R},-\boldsymbol{p})^{T}
\end{pmatrix}.
\end{equation}
Here, for simplicity, we have chosen $\Delta(\boldsymbol{R})$ to be real-valued. From a matrix inversion, we find for the off-diagonal components, 
\begin{equation}
F_{\uparrow\uparrow}(i\omega_{n})
=
\frac{
2\Delta\,\alpha_{\boldsymbol{p}}(\xi-b_{z})
}
{
(\omega_{n}^{2}+E_{+}^{2})(\omega_{n}^{2}+E_{-}^{2})
},
\quad
F_{\downarrow\downarrow}(i\omega_{n})
=
-
\frac{
2\Delta\,\alpha_{\boldsymbol{p}}(\xi+b_{z})
}
{
(\omega_{n}^{2}+E_{+}^{2})(\omega_{n}^{2}+E_{-}^{2})
}.
\end{equation}
Here, we have dropped the $(\boldsymbol{R},\boldsymbol{p})$ arguments from the respective expressions for notational brevity. The local quasiparticle energies are given by,
\begin{equation}
E_{\pm}^{2}
=
\Delta^{2}+\xi^{2}+\alpha_{\boldsymbol{p}}^{2}+b_{z}^{2}
\pm
2\sqrt{
\xi^{2}(\alpha_{\boldsymbol{p}}^{2}+b_{z}^{2})
+
\Delta^{2} b_{z}^{2}
}.
\end{equation}
We can now compute the equal-time pairing correlation by summing over the Matsubara frequencies,
\begin{equation}
\mathcal{F}_{\sigma\sigma}(\boldsymbol{R},\boldsymbol{p})
=
T\sum_{\omega_{n}}
F_{\sigma\sigma}(\boldsymbol{R},\boldsymbol{p};i\omega_{n}).
\end{equation}
We find that,
\begin{equation}
\mathcal{F}_{\sigma\sigma}
=
\alpha_{\boldsymbol{p}}
\left(
\sigma\xi-b_{z}
\right)
\frac{\Delta}{E_{+}^{2}-E_{-}^{2}}
\left[
\frac{\tanh(E_{-}/2T)}{E_{-}}
-
\frac{\tanh(E_{+}/2T)}{E_{+}}
\right].
\end{equation}
The result in the main text is the zero-temperature limit of this expression. 
Here, we have used that, 
\begin{equation}
\frac{1}{(\omega_{n}^{2}+E_{+}^{2})(\omega_{n}^{2}+E_{-}^{2})}
=
\frac{1}{E_{+}^{2}-E_{-}^{2}}
\left(
\frac{1}{\omega_{n}^{2}+E_{-}^{2}}
-
\frac{1}{\omega_{n}^{2}+E_{+}^{2}}
\right),
\end{equation}
and the Matsubara sum, 
\begin{equation}
T\sum_{n} \frac{1}{\omega_{n}^{2}+E^{2}}
=
\frac{1}{2E}\tanh\left(\frac{E}{2T}\right).
\end{equation}

\section{Supercurrent-induced torques}

\subsection{Free energy density}
In this first subsection, we provide more details on obtaining the free energy density of the BdG Hamiltonian with a
finite Cooper-pair momentum. 
\\
\\
We begin by redisplaying the BdG Hamiltonian with a finite Cooper-pair momentum, 
\begin{equation}
\mathcal{H}_{\text{BdG}}(\boldsymbol{R},\boldsymbol{p};\boldsymbol{Q})
=
\begin{pmatrix}
h_{\text{proj}}(\boldsymbol{R},\boldsymbol{p}+\hbar\boldsymbol{Q}/2)
&
\Delta(\boldsymbol{R})(-i\sigma_{y})
\\
-\Delta(\boldsymbol{R})(-i\sigma_{y})
&
-\left[
h_{\text{proj}}(\boldsymbol{R},-\boldsymbol{p}+\hbar\boldsymbol{Q}/2)
\right]^{T}
\end{pmatrix},
\end{equation}
and the free energy,
\begin{equation}
\Omega[\phi,\boldsymbol{Q}]
=
-\frac{T}{2}
\sum_{\omega_{n}}
\text{Tr}
\ln
\left[
i\omega_{n}
-
\mathcal{H}_{\text{BdG}}(\boldsymbol{Q})
\right].
\end{equation}
For the subsequent calculations, it will be useful to write the free energy more explicitly in terms of the eigenbalues of the BdG Hamiltonian. We will denote these eigenvalues by $E_{a}(\boldsymbol{R},\boldsymbol{p};\boldsymbol{Q})$ with $a=1,\cdots,4$. We can note that (up to an energy-independent constant), 
\begin{equation}
-\frac{T}{2} \sum_{\omega_{n}} \ln \left[ i\omega_{n}-E \right] = \frac{1}{2}g(E)    
\quad
\text{with}
\quad
g'(E)=f(E)
\quad
\text{and}
\quad
f(E)=\frac{1}{e^{E/T}+1}.
\end{equation}
As a result, the change in the free energy due to the finite Cooper-pair momentum can be written as, 
\begin{equation}
\Omega_{Q}
\equiv
\Omega[\phi,\boldsymbol{Q}]-\Omega[\phi,\boldsymbol{0}]
=
\frac{1}{2}
\int d^{2}R\,
\int
\frac{d^{2}p}{(2\pi\hbar)^{2}}
\sum_{a=1}^{4}
\left[
g(E_{a}(\boldsymbol{R},\boldsymbol{p};\boldsymbol{Q}))
-
g(E_{a}(\boldsymbol{R},\boldsymbol{p};\boldsymbol{0}))
\right]
\end{equation}
We remark that this change in the free energy should, in principle, also comprise a contribution from the double of the degrees of freedom
in the Nambu representation. However, this contribution is texture-independent. It will thus not be of relevance for our subsequent
considerations. It reads, 
\begin{equation}
\frac{1}{2}
\int d^{2}R\,
\int
\frac{d^{2}p}{(2\pi\hbar)^{2}}
\text{tr}
\left[
h_{\text{proj}}(\boldsymbol{R},\boldsymbol{p}+\boldsymbol{q})
-
h_{\text{proj}}(\boldsymbol{R},\boldsymbol{p})
\right]
=
\int d^{2}R\,
\int
\frac{d^{2}p}{(2\pi\hbar)^{2}}
\left[
\xi(\boldsymbol{R},\boldsymbol{p}+\boldsymbol{q})
-
\xi(\boldsymbol{R},\boldsymbol{p})
\right],
\end{equation}
where the trace is over the effective spin degrees of freedoms and we used that $\sigma_{x}$ and $\sigma_{z}$ are traceless.

\subsection{Superfluid stiffness}
In this second subsection, we provide more details on obtaining an expression for the superfluid stiffness. 
\\
\\
The superfluid stiffness, $D^{ij}(\boldsymbol{R})$, is defined through the expansion of the free energy, 
\begin{equation}
\Omega_{Q}
=
\int d^{2}R\,
Q_{i}J^{i}(\boldsymbol{R})
+
\frac{1}{2}
\int d^{2}R\,
Q_{i}Q_{j}D^{ij}(\boldsymbol{R})
+
\dots.
\end{equation}
To obtain an explicit expression for it, we note the expansion, 
\begin{equation}
g(E_{a}(\boldsymbol{p};\boldsymbol{Q}))
=
g(E_{a})
+
Q_{i} f(E_{a})v_{a}^{i}
+
\frac{1}{2}Q_{i}Q_{j}
\left[
f'(E_{a})v_{a}^{i}v_{a}^{j}
+
f(E_{a})m_{a}^{ij}
\right]
+
\dots,
\end{equation}
where we have adopted the short-hand notations, 
\begin{equation}
E_{a}
\equiv
E_{a}(\boldsymbol{p};\boldsymbol{0}),
\quad
v_{a}^{i}
\equiv
\left.
\frac{\partial E_{a}(\boldsymbol{p};\boldsymbol{Q})}
{\partial Q_{i}}
\right|_{\boldsymbol{Q}=0},
\quad
m_{a}^{ij}
\equiv
\left.
\frac{\partial^{2} E_{a}(\boldsymbol{p};\boldsymbol{Q})}
{\partial Q_{i}\partial Q_{j}}
\right|_{\boldsymbol{Q}=0}.
\end{equation}
A direct comparsion then gives the expression, 
\begin{equation}
D^{ij}
=
\frac{1}{2}
\int
\frac{d^{2}p}{(2\pi\hbar)^{2}}
\sum_{a=1}^{4}
\left[
f'(E_{a})v_{a}^{i}v_{a}^{j}
+
f(E_{a})m_{a}^{ij}
\right].
\end{equation}

\subsection{Gradient expansion of the superfluid stiffness}
In this third subsection, we provide more details on the gradient expansion of the superfluid stiffness. 
Specifically, up to second order in gradients of the texture angle, we have, 
\begin{equation}
D^{ij}
=
\bar{D}^{ij}
+
\varrho_{0}\bar{C}_{0}^{ij}V_{0}
+
\varrho_{z}\bar{C}_{z}^{ij}V_{z}
+
\frac{1}{2}
\bar{A}^{ij;mn}\alpha_{m}\alpha_{n}
+
\mathcal{O}\left[(\partial_{i}\phi)^{4}\right].
\end{equation}
Here, we have defined, 
\begin{equation}
\bar{D}^{ij}
=
\overline{D^{ij}},
\quad
\bar{C}_{0}^{ij}
=
\overline{
\frac{\partial D^{ij}}{\partial(\varrho_{0}V_{0})}
},
\quad
\bar{C}_{z}^{ij}
=
\overline{
\frac{\partial D^{ij}}{\partial(\varrho_{z}V_{z})}
},
\quad
\bar{A}^{ij;mn}
=
\overline{
\frac{\partial^{2}D^{ij}}{\partial\alpha_{m}\partial\alpha_{n}}
}.
\end{equation}
The overline should be understood as evaluating the derivative first and, subsequently, evaluating the expression in the uniform limit,
\begin{equation}
V_{0}=V_{z}=\alpha_{i}=0.    
\end{equation}
We note that there is no term linear in $\alpha_{m}$ in the expansion. 
Changing $\alpha_{m}\to-\alpha_{m}$, or equivalently $\alpha_{\boldsymbol{p}}\sigma_{x}\to-\alpha_{\boldsymbol{p}}\sigma_{x}$, is generated by a $\pi$-rotation around the $z$-axis in spin space. 
Since this is a basis transformation, the spectrum and the superfluid stiffness are invariant under $\alpha_{m}\to-\alpha_{m}$. 

\subsection{Quasiparticle spectrum in limiting cases}
In this fourth subsection, we provide the quasiparticle spectrum of the proximitized planar domain wall in limiting cases that will be of relevance for our subsequent discussions. 
\\
\\
First, in the absence of the emergent spin-orbit coupling, $\alpha_{\boldsymbol{p}}=0$, the quasiparticle spectrum is given by,
\begin{equation}
\begin{split}
E_{s\tau}(\boldsymbol{p};\boldsymbol{Q})
&=
\frac{
\xi(\boldsymbol{p}+\hbar\boldsymbol{Q}/2)
-
\xi(\boldsymbol{p}-\hbar\boldsymbol{Q}/2)
}{2}
+
\frac{s}{2}
\left[
b_{z}(\boldsymbol{p}+\hbar\boldsymbol{Q}/2)
+
b_{z}(\boldsymbol{p}-\hbar\boldsymbol{Q}/2)
\right]
\\
&\quad
+
\tau
\sqrt{
\Delta(\boldsymbol{R})^{2}
+
\frac{1}{4}
\left[
\xi(\boldsymbol{p}+\hbar\boldsymbol{Q}/2)
+
\xi(\boldsymbol{p}-\hbar\boldsymbol{Q}/2)
+
s\left(
b_{z}(\boldsymbol{p}+\hbar\boldsymbol{Q}/2)
-
b_{z}(\boldsymbol{p}-\hbar\boldsymbol{Q}/2)
\right)
\right]^{2}
}.
\end{split}
\end{equation}
Second, if also the Cooper-pair momentum vanishes, $\boldsymbol{Q}=0$, then the quasiparticle spectrum is given by,
\begin{equation}
E_{s\tau}(\boldsymbol{p})
\equiv
E_{s\tau}(\boldsymbol{p};\boldsymbol{0})
=
s b_{z}(\boldsymbol{p})
+
\tau
\sqrt{\xi(\boldsymbol{p})^{2}+\Delta^{2}}.
\end{equation}
Third, if we adopt also the uniform limit where also $V_{0}=V_{z}=0$, then the quasiparticle spectrum takes on the form, 
\begin{equation}
\bar{E}_{s\tau}(\boldsymbol{p})
=
s\varrho_{z}\eta^{mn}p_{m}p_{n}
+
\tau
\sqrt{
(\varrho_{0}p^{2}-\mu)^{2}+\Delta^{2}
}.
\end{equation}

\subsection{Superfluid stiffness coefficient}
In this fifth subsection, we provide more details on evaluating the coefficient $\bar C_z^{ij}$. 
\\
\\
First, we write the superfluid stiffness in the form, 
\begin{equation}
D^{ij}
=
\frac{1}{2}
\int
\frac{d^2p}{(2\pi\hbar)^2}
\sum_{a=1}^{4}
\mathcal D_a^{ij}    
\quad
\text{with}
\quad 
\mathcal D_a^{ij}
\equiv
f'(E_a)v_a^iv_a^j
+
f(E_a)m_a^{ij}.
\end{equation}
The coefficient $\bar C_z^{ij}$ is then given by,
\begin{equation}
\bar C_z^{ij}
=
\frac{1}{2}
\int
\frac{d^2p}{(2\pi\hbar)^2}
\sum_{a=1}^{4}
\overline{
\frac{\partial\mathcal D_a^{ij}}
{\partial(\varrho_zV_z)}
}.
\end{equation}
Second, we apply the chain rule to find, 
\begin{equation}
\frac{\partial\mathcal D_a^{ij}}{\partial(\varrho_zV_z)}
=
\frac{\partial E_a}{\partial(\varrho_zV_z)}
\left[
f''(E_a)v_a^iv_a^j
+
f'(E_a)m_a^{ij}
\right]
+
f'(E_a)
\left[
\frac{\partial v_a^i}{\partial(\varrho_zV_z)}v_a^j
+
v_a^i
\frac{\partial v_a^j}{\partial(\varrho_zV_z)}
\right]
+
f(E_a)
\frac{\partial m_a^{ij}}{\partial(\varrho_zV_z)}.
\end{equation}
We will be interested in this expression in the uniform limit. When $\alpha_{\boldsymbol p}=0$, the geometric Zeeman field, $V_z$, enters the quasiparticle energies, $E_{s\tau}(\boldsymbol p;\boldsymbol Q)$, only as a $\boldsymbol{Q}$-independent shift. As a result, 
\begin{equation}
\overline{
\frac{\partial v_a^i}{\partial(\varrho_zV_z)}
}
=
0
\quad
\text{and}
\quad
\overline{
\frac{\partial m_a^{ij}}{\partial(\varrho_zV_z)}
}
=
0.
\end{equation}
We then find the following expressions for the superfluid stiffness coefficient,
\begin{equation}
\bar C_z^{ij}
=
\frac{1}{2}
\int
\frac{d^2p}{(2\pi\hbar)^2}
\sum_{a=1}^{4}
\left(
\overline{
\frac{\partial E_a}{\partial(\varrho_zV_z)}
}
\right)
\left[
f''(\bar E_a)
\bar v_a^i\bar v_a^j
+
f'(\bar E_a)
\bar m_a^{ij}
\right].
\end{equation}
In the uniform limit, we can label the quasiparticle energies by $a=(s,\tau)$. In this case, the above expressions simplifies to, 
\begin{equation}
\bar C_z^{ij}
=
\frac{1}{2}
\int
\frac{d^2p}{(2\pi\hbar)^2}
\sum_{s,\tau=\pm1}
s
\left[
f''(\bar E_{s\tau})
\bar v_{s\tau}^i\bar v_{s\tau}^j
+
f'(\bar E_{s\tau})
\bar m_{s\tau}^{ij}
\right].
\end{equation}
We now want to derive the tensorial structure of $\bar C_z^{ij}$. We therefore focus on the relevant limit when $V_0=\alpha_{\boldsymbol{p}}=0$. 
In this case, the system is invariant under an exchange of coordinates, $x\leftrightarrow y$, combined with a spin-flip operation. More precisely, at the level of the normal-state Hamiltonian, 
\begin{equation}
U_x
h_{\text{proj}}(
X\boldsymbol p;-V_z)
U_x^\dagger
=
h_{\text{proj}}(\boldsymbol p;V_z)
\quad
\text{with}
\quad
X
\equiv
\begin{pmatrix}
0&1\\
1&0
\end{pmatrix}
\quad
\text{and}
\quad
U_x=-i\sigma_x.
\end{equation}
At the level of the BdG Hamiltonian, 
\begin{equation}
\mathcal U_x
\mathcal H_{\text{BdG}}
(
X\boldsymbol p;
X\boldsymbol Q,
-V_z
)
\mathcal U_x^\dagger
=
\mathcal H_{\text{BdG}}
(
\boldsymbol p;
\boldsymbol Q,
V_z
)
\quad
\text{with}
\quad
\mathcal U_x
=
\begin{pmatrix}
U_x&0\\
0&U_x^*
\end{pmatrix}.
\end{equation}
Because the trace-log is invariant under unitary transformations and the momentum integration is unchanged under
$\boldsymbol p\to X\boldsymbol p$, the free energy satisfies,
\begin{equation}
\Omega_Q(V_z,\boldsymbol Q)
=
\Omega_Q(-V_z,X\boldsymbol Q).
\end{equation}
To fulfill this requirement, the superfluid stiffness needs to satisfy,
\begin{equation}
D^{ij}(V_z)
=
X^{i}{}_{k}
X^{j}{}_{l}
D^{kl}(-V_z)
\end{equation}
and the $V_z$ coefficient in the gradient expansion of the superfluid stiffness needs to satisfy,
\begin{equation}
\bar C_z^{ij}
=
-
X^{i}_{k}
X^{j}_{l}
\bar C_z^{kl}.
\end{equation}
This condition constrains the $V_z$ to the form, 
\begin{equation}
\bar C_z^{ij}
=
\bar C_z\eta^{ij}
\quad
\text{or, equivalently,}
\quad
\bar C_z
=
\frac{1}{2}
\eta_{ij}\bar C_z^{ij}.
\end{equation}
In particular, we note that the contribution to the superfluid stiffness that is $\propto V_z$ can be written as,
\begin{equation}
\begin{split}
D_Q^{V_z}
&\equiv
\varrho_z V_z\hat Q_i\hat Q_j\bar C_z^{ij}
\\
&=
\varrho_z V_z
\bar{C}_z(\hat Q_x^2-\hat Q_y^2)
\\
&=
\varrho_z V_z
\bar{C}_z\cos 2\chi_Q
\end{split}    
\end{equation}
with the parameterization 
$\hat{\boldsymbol Q}
=
(\cos\chi_Q,\sin\chi_Q)$.

\subsection{Torque contributions}
In this sixth subsection, we provide more details on the derivation of the contributions to the supercurrent-induced torque. 
\\
\\
We will focus on the planar radial texture with Néel vector 
$\boldsymbol{n}(\boldsymbol{R})
=
(\cos\phi(\boldsymbol{R}),\sin\phi(\boldsymbol{R}),0)$.
Since the Néel vector varies only within the $x$-$y$-plane, the only component of the supercurrent-induced torque is in the $z$-direction and given by, 
\begin{equation}
\tau_{Q}(\boldsymbol{R})
=
-\hat{\boldsymbol{z}}\cdot
\left(
\boldsymbol{n}
\times
\frac{\delta\Omega_{Q}}{\delta\boldsymbol{n}}
\right).
\end{equation}
To evaluate this torque component, we consider the variation of the supercurrent-induced change of the free energy density with respect to $\boldsymbol{n}(\boldsymbol{R})$, or, equivalently, 
$\phi(\boldsymbol{R})$. We find that,
\begin{equation}
\begin{split}
\int d^{2}R\,
\frac{\delta\Omega_{Q}}{\delta\phi}
\delta\phi=\delta\Omega
&=
\int d^{2}R\,
\frac{\delta\Omega_{Q}}{\delta\boldsymbol{n}}
\cdot
\delta\boldsymbol{n}  
\\
&
=
\int d^{2}R\,
\frac{\delta\Omega_{Q}}{\delta\boldsymbol{n}}
\cdot
(\hat{\boldsymbol{z}}\times\boldsymbol{n})\,
\delta\phi 
\\
&=
\int d^{2}R\,
\hat{\boldsymbol{z}}\cdot
\left(
\boldsymbol{n}
\times
\frac{\delta\Omega_{Q}}{\delta\boldsymbol{n}}
\right)
\delta\phi.
\end{split}    
\end{equation}
We thus conclude that, 
\begin{equation}
\tau_{Q}(\boldsymbol{R})
=
-\frac{\delta\Omega_{Q}}{\delta\phi}. 
\end{equation}
As a next step, we return to the gradient expansion of the superfluid stiffness, 
\begin{equation}
D_{Q}
=
\bar{D}_{Q}
+
\varrho_{0}\bar{C}_{0,Q}V_{0}
+
\varrho_{z}\bar{C}_{z,Q}V_{z}
+
\frac{1}{2}\bar{A}_{Q}^{mn}\alpha_{m}\alpha_{n}
+
\mathcal{O}\!\left((\partial_{i}\phi)^{4}\right),
\end{equation}
with 
$
\bar{D}_{Q}=\hat{Q}_{i}\hat{Q}_{j}\bar{D}^{ij}
$,
$
\bar{C}_{0,Q}=\hat{Q}_{i}\hat{Q}_{j}\bar{C}_{0}^{ij}
$,
$
\bar{C}_{z,Q}=\hat{Q}_{i}\hat{Q}_{j}\bar{C}_{z}^{ij}
$,
and
$
\bar{A}_{Q}^{mn}=\hat{Q}_{i}\hat{Q}_{j}\bar{A}^{ij;mn}
$.
We now recall that for our $d$-wave radial domain wall texture, we have, 
\begin{equation}
V_{0}
=
\frac{\hbar^{2}}{4}
\delta^{ij}\partial_{i}\phi\,\partial_{j}\phi,
\quad
V_{z}
=
\frac{\hbar^{2}}{4}
\eta^{ij}\partial_{i}\phi\,\partial_{j}\phi,
\quad
\alpha_{i}
=
\hbar t_{3}\partial_{i}\phi    
\end{equation}
with the variations, 
\begin{equation}
\delta V_{0}
=
\frac{\hbar^{2}}{2}
\delta^{ij}\partial_{i}\phi\,\partial_{j}\delta\phi,
\quad
\delta V_{z}
=
\frac{\hbar^{2}}{2}
\eta^{ij}\partial_{i}\phi\,\partial_{j}\delta\phi,
\quad
\delta\alpha_{i}
=
\hbar t_{3}\partial_{i}\delta\phi.
\end{equation}
We can now use these expressions for computing the variation of the free energy, 
\begin{equation}
\begin{split}
\delta\Omega_{Q}
&\approx
\frac{Q^{2}}{2}
\int d^{2}R\,
\left[
\varrho_{0}\bar{C}_{0,Q}\delta V_{0}
+
\varrho_{z}\bar{C}_{z,Q}\delta V_{z}
+
\bar{A}_{Q}^{mn}\alpha_{n}\delta\alpha_{m}
\right]
\\
&=
\frac{Q^{2}}{2}
\int d^{2}R\,
T_{Q}^{j}\,\partial_{j}\delta\phi \\
&=
-\frac{Q^{2}}{2}
\int d^{2}R\,
(\partial_{j} T_{Q}^{j})\delta\phi
\end{split}    
\end{equation}
In the last equality, we have integrated-by-parts and excluded the boundary term due to periodic boundary conditions in $\phi$. Moreover, we have defined,
\begin{equation}
T_{Q}^{j}
=
\frac{\hbar^{2}\varrho_{0}}{2}
\bar{C}_{0,Q}\delta^{ij}\partial_{i}\phi
+
\frac{\hbar^{2}\varrho_{z}}{2}
\bar{C}_{z,Q}\eta^{ij}\partial_{i}\phi
+
\hbar^{2}t_{3}^{2}
\bar{A}_{Q}^{jn}\partial_{n}\phi.
\end{equation}
We can now read off the torque as, 
\begin{equation}
\tau_{Q}
=
\tau_{Q}^{V_{0}}
+
\tau_{Q}^{V_{z}}
+
\tau_{Q}^{\text{SOC}} .
\end{equation}
with
\begin{equation}
\tau_{Q}^{V_{0}}
=
\frac{Q^{2}}{2}
\partial_{j}
\left[
\frac{\hbar^{2}\varrho_{0}}{2}
\bar{C}_{0,Q}
\delta^{ij}\partial_{i}\phi
\right],
\quad
\tau_{Q}^{V_{z}}
=
\frac{Q^{2}}{2}
\partial_{j}
\left[
\frac{\hbar^{2}\varrho_{z}}{2}
\bar{C}_{z,Q}
\eta^{ij}\partial_{i}\phi
\right],
\quad
\tau_{Q}^{\alpha}
=
\frac{Q^{2}}{2}
\partial_{j}
\left[
\hbar^{2}t_{3}^{2}
\bar{A}_{Q}^{jn}\partial_{n}\phi
\right].
\end{equation}

\newpage

\section{Domain wall deformation}
In this section, we will provide details on the deformation of the planar radial domain wall with Néel vector $\boldsymbol{n}(\boldsymbol{r})=(\cos\phi_{0}(r),\sin\phi_{0}(r),0)$ due to the altermagnetic torque. Here, $\phi_{0}(r)=(\pi/2)
\tanh((r-R_{0})/w)$ is the texture angle in the absence of the torque.
\\
\\
Suppose that the altermagnetic torque,
\begin{equation}
\tau^{V_{z}}_{Q}(\boldsymbol{r})
=
T_{Q}\cos2\chi
\left[
\phi_{0}''(r)-\frac{\phi_{0}'(r)}{r}
\right]
\quad
\text{with}
\quad
T_{Q}
=
\frac{\hbar^{2}\varrho_{z} Q^{2}}{4}
\bar{C}_{z}
\cos2\chi_{Q},
\end{equation}
gives rise to a small deformation of the Néel texture, $\phi_{0}(r)\to\phi_{0}(r)+\delta\phi(\boldsymbol{r})$. In this case the work done by torque is given by, 
\begin{equation}
\delta W_{Q}
=
\int d^{2}r
\,
\tau^{V_{z}}_{Q}(\boldsymbol{r})
\delta\phi(\boldsymbol{r}).
\end{equation}
Suppose now that this deformation of the Néel texture is due to an angular-dependent change of the domain wall radius,
\begin{equation}
R(\chi)=R_{0}+u(\chi).    
\end{equation}
In this case the angle in the texture angle changes to, 
\begin{equation}
\begin{split}
\phi(r,\chi)
&\equiv
\phi_{0}(r-u(\chi))
\\
&\approx
\phi_{0}(r)-u(\chi)\phi_{0}'(r).
\end{split}
\end{equation}
Hence, for a small radial displacement $\delta u(\chi)$, we have a change of the texture angle, 
\begin{equation}
\delta\phi(r,\chi)
=
-\phi_{0}'(r)\delta u(\chi).
\end{equation}
The resulting work done on the texture is given by, 
\begin{equation}
\begin{split}
\delta W_{Q}
&=
\int r \, 
dr \, 
d\chi \,
\tau^{V_{z}}_{Q}(r,\chi)
[-\phi_{0}'(r)\delta u(\chi)]
\\
&=
-T_{Q}
\int d\chi
\,
\cos2\chi
\,
\delta u(\chi)
\int_{0}^{\infty} r
\,
dr
\,
\phi_{0}'(r)
\left[
\phi_{0}''(r)-\frac{\phi_{0}'(r)}{r}
\right]
\\
&\equiv
\frac{T_{Q}\pi^{2}}{2w}
\int d\chi \,
\cos2\chi 
\,
\delta u(\chi)
\end{split}    
\end{equation}
where we have evaluated the radial integral in the limit when $R_{0}\gg w$,
\begin{equation}
\int_{0}^{\infty} r
\,
dr
\,
\phi_{0}'(r)
\left[
\phi_{0}''(r)-\frac{\phi_{0}'(r)}{r}
\right]
\approx 
-
 \frac{\pi^{2}}{2w}
\end{equation}
It is now useful to perform a Fourier expansion $\delta u(\chi)$,
\begin{equation}
\delta u(\chi)
=
\delta u_{0}
+
\delta u_{1}\cos\chi
+
\delta u'_{1}\sin\chi
+
\delta u_{2}\cos2\chi
+
\delta u'_{2}\sin2\chi
+\dots
\end{equation}
All shown contributions except the one $\propto \delta u_{2}$ vanish upon insertion into the expression for $\delta W_{Q}$.
As a result, 
\begin{equation}
\begin{split}
\delta W_{Q}
=
F_{Q}
\delta u_{2}
\quad
\text{with}
\quad
F_{Q}
=
\frac{\pi^{3}}{8w}
\hbar^{2}\varrho_{z} Q^{2}
\bar{C}_{z}
\cos2\chi_{Q}
\end{split}    
\end{equation}
where $F_{Q}$ is a generalized force that acts on the coordinate $u_{2}$. 
\\
\\
In equilibrium, we assume that this force is approximately balanced by a restoring force that is linear in $u_{2}$, $F_{\text{rest}} = - k u_{2}$. In equilibrium, when $F_{Q} + F_{\text{rest}}=0$, we find, 
\begin{equation}
u_{2} 
= 
\frac{\pi^{3}}{8wk}
\hbar^{2}\varrho_{z} Q^{2}
\bar{C}_{z}
\cos2\chi_{Q}.
\end{equation}
In this situation, the parameterization of the domain wall is given by, 
\begin{equation}
R(\chi)=R_{0}+u_{2}\cos2\chi,    
\end{equation}
which shows that the domain wall takes on an elliptical shape.

\section{Simulation details}

\subsection{Simulation parameters for Fig.~2}
In Fig.~2 of the main text, we plot the local quasiparticle spectrum for the radial domain wall given by,
\begin{equation}
E_{\pm}^{2}
=
\Delta^{2}+\xi^{2}+\alpha_{\boldsymbol{p}}^{2}+b_{z}^{2}
\pm
2\sqrt{\xi^{2}(\alpha_{\boldsymbol{p}}^{2}+b_{z}^{2})+\Delta^{2}b_{z}^{2}}.
\end{equation}
The relevant quantities entering this expression are, 
\begin{equation}
\begin{split}
\xi
&=
\varrho_{0} p^{2}
+
\frac{\varrho_{0}}{4}\phi'(R)^{2}
-
\mu,
\hspace{2.5cm}
\alpha_{\boldsymbol{p}}
=
\varrho_{3}\phi'(R)
\bigl(p_{x}\cos\chi+p_{y}\sin\chi\bigr),
\\
b_{z}
&=
\varrho_{z}(p_{x}^{2}-p_{y}^{2})
+
\frac{\varrho_{z}}{4}\phi'(R)^{2}\cos2\chi,
\hspace{0.95cm}
\phi'(r)
=
\frac{\pi}{2w}\,\text{sech}^{2}\!\left(\frac{r-R_{0}}{w}\right).
\end{split}    
\end{equation}
The simulation parameters are,
\begin{table}[h!]
\centering
\renewcommand{\arraystretch}{1.15}
\begin{tabular}{@{}cccccccccc@{}}
$\varrho_{0}$ & $\varrho_{z}$ & $\varrho_{3}$ & $\mu$ & $\Delta$ & $R_{0}$ & $w$  \\
\midrule
$1.00$ & $0.40$ & $0.70$ & $1.70$ & $0.25$ & $6.0$ & $1.65$  \\
\bottomrule
\end{tabular}
\caption{
\textbf{Simulation parameters for Fig.\,2.}
}
\end{table}\\
We remark that in all simulations, we set $\hbar=1$ and use the momentum scale $p_{0}=\sqrt{\mu/\varrho_{0}}$.

\subsection{Simulation parameters for Fig.~3}
In Fig.~3 of the main text, we plot the equal-spin correlation function, 
\begin{equation}
\mathcal{F}_{\uparrow\uparrow}
=
\alpha_{\boldsymbol{p}}(\xi-b_{z})
\frac{\Delta}{E_{+}^{2}-E_{-}^{2}}
\left[
\frac{\tanh(E_{-}/2T)}{E_{-}}
-
\frac{\tanh(E_{+}/2T)}{E_{+}}
\right].
\end{equation}
The simulation parameters are, 
\begin{table}[h!]
\centering
\renewcommand{\arraystretch}{1.15}
\begin{tabular}{@{}ccccccccc@{}}
$\varrho_{0}$ & $\varrho_{z}$ & $\varrho_{3}$ & $\mu$ & $\Delta$ & $T$ & $R_{0}$ & $w$ \\
\midrule
$1.00$ & $0.40$ & $0.70$ & $1.70$ & $0.25$ & $0.06$ & $6.0$ & $1.65$  \\
\bottomrule
\end{tabular}
\caption{
\textbf{Simulation parameters for Fig.\,3.}
}
\end{table}

\subsection{Simulation parameters for Fig.~4}
In Fig.~4 of the main text, we plot the triplet intensity, 
\begin{equation}
I^{\sigma\sigma}_{t}(\boldsymbol{R})
=
\left|
\int_{|\boldsymbol{p}|\leq p_{c}}
\frac{d^{2}p}{(2\pi\hbar)^{2}}\,
\boldsymbol{p}\,
\mathcal{F}_{\sigma\sigma}(\boldsymbol{R},\boldsymbol{p})
\right|^{2}. 
\end{equation}
The simulation parameters are, 
\begin{table}[h!]
\centering
\renewcommand{\arraystretch}{1.15}
\begin{tabular}{@{}ccccccccccc@{}}
$\varrho_{0}$ & $\varrho_{z}$ & $\varrho_{3}$ & $\mu$ & $\Delta$ & $T$ & $R_{0}$ & $w$ & $p_{c}$\\
\midrule
$1.00$ & $0.70$ & $0.90$ & $1.70$ & $0.25$ & $0.06$ & $6.0$ & $1.10$ & $2.15p_{0}$  \\
\bottomrule
\end{tabular}
\caption{
\textbf{Simulation parameters for Fig.\,4.}
}
\end{table}

\subsection{Simulation details for Fig.~5}
In Fig.~5 of the main text, we plot the altermagnetic torque for a given $\hat{\boldsymbol{Q}}=(\cos\chi_{Q},\sin\chi_{Q})$. It is given by, 
\begin{equation}
\tau_{Q}^{V_{z}}(r,\chi)
\propto
\cos 2\chi_{Q}\,\cos 2\chi
\left[
\phi''(r)-\frac{\phi'(r)}{r}
\right].
\end{equation}
We plot the results in terms of the dimensionless radial coordinate, $r/R_{0}$. We further choose $R_{0}/w=7.5$. 

\subsection{Simulation details for Fig.~6}
In Fig.~6 of the main text, we evaluate the torque coefficient,
\begin{equation}
\bar{C}_{z}
=
\frac{\eta_{ij}}{4}
\int
\frac{d^{2}p}{(2\pi\hbar)^{2}}
\sum_{s,\tau=\pm1}
s
\left[
f''(\bar{E}_{s\tau})
\bar{v}_{s\tau}^{,i}\bar{v}_{s\tau}^{,j}
+
f'(\bar{E}_{s\tau})
\bar{m}_{s\tau}^{,ij}
\right].
\end{equation}
In our simulations, we set $\hbar=k_{B}=1$.
For the uniform quasiparticle levels, we take,
\begin{equation}
\bar{E}_{s\tau}
=
s\varrho_{z}p^{2}\cos 2\theta
+
\tau
\sqrt{\xi^{2}+\Delta(T)^{2}}
\quad
\text{with}
\quad
\xi=\varrho_{0}p^{2}-\mu.
\end{equation}
For the temperature dependence of the superconducting gap, we choose, 
\begin{equation}
\Delta(T)
=
\Delta_{0}
\tanh\left(
1.74
\sqrt{\frac{T_{c}}{T}-1}
\right)
\quad
\text{with}
\quad
\Delta_{0}=1.76T_{c}.
\end{equation}
The additional terms in the expression for the torque coefficient are given by, 
\begin{equation}
\begin{split}
\eta_{ij}
\bar{v}_{s\tau}^{i}\bar{v}_{s\tau}^{j}
&=
\varrho_{0}^{2}p^{2}\cos2\theta
+
2s\tau\varrho_{0}\varrho_{z}
\frac{\xi}{\sqrt{\xi^{2}+\Delta^{2}}}
p^{2}
+
\varrho_{z}^{2}
\frac{\xi^{2}}{\xi^{2}+\Delta^{2}}
p^{2}\cos2\theta,
\\
\eta_{ij}
\bar{m}_{s\tau}^{ij}
&=
s\varrho_{z}
+
\tau\varrho_{z}^{2}
\frac{\Delta^{2}}
{\left(\xi^{2}+\Delta^{2}\right)^{3/2}}
p^{2}\cos2\theta,
\end{split}
\end{equation}
where we have parameterized the momentum as $p_{x}=p\cos\theta$ and $p_{y}=p\sin\theta$. We then perform a change of variables, $p^{2}=(\xi+\mu)/\varrho_{0}$, so that the integration measures changes to, 
\begin{equation}
\int\frac{d^{2}p}{(2\pi)^{2}}
=
\frac{1}{2\varrho_{0}(2\pi)^{2}}
\int_{-\mu}^{\xi_{\text{cutoff}}}d\xi
\int_{0}^{2\pi}d\theta.
\end{equation}
The simulation parameters are, 
\begin{table}[h!]
\centering
\renewcommand{\arraystretch}{1.15}
\begin{tabular}{@{}cccccccc@{}}
$T_{c}$ & $\Delta_{0}$ & $\varrho_{0}$ & $\mu$ & $\xi_{\text{cutoff}}$ \\
\midrule
$1.0$ & $1.76$ & $1.0$ & $12.0$ & $72.0$  \\
\bottomrule
\end{tabular}
\caption{
\textbf{Simulation parameters for Fig.\,6.}
}
\end{table}

\end{document}